\documentclass[aps,prl,notitlepage,12pt,superscriptaddress]{revtex4-1}
\usepackage{amsmath,amssymb}
\usepackage{url}
\usepackage{graphicx}
\usepackage{xcolor}
\usepackage{soul}
\setstcolor{purple}

\newcommand{\nocontentsline}[3]{}
\newcommand{\tocless}[2]{\bgroup\let\addcontentsline=\nocontentsline#1{#2}\egroup}

\begin{document}
\title{Modeling the growth of organisms validates a general relation\\ between metabolic costs and natural selection}

\author{Efe Ilker}
\affiliation{Physico-Chimie Curie UMR 168, Institut Curie, PSL Research University, 26 rue d'Ulm, 75248 Paris Cedex 05, France}
\affiliation{Department of Physics, Case Western Reserve University, Cleveland OH 44106}
\author{Michael Hinczewski}
\affiliation{Department of Physics, Case Western Reserve University, Cleveland OH 44106}

\begin{abstract}
  Metabolism and evolution are closely connected: if a mutation incurs extra energetic costs for an organism, there is a baseline selective disadvantage that may or may not be compensated for by other adaptive effects. A long-standing, but to date unproven, hypothesis is that this disadvantage is equal to the fractional cost relative to the total resting metabolic expenditure. This hypothesis has found a recent resurgence as a powerful tool for quantitatively understanding the strength of selection among different classes of organisms. Our work explores the validity of the hypothesis from first principles through a generalized metabolic growth model, versions of which have been successful in describing organismal growth from single cells to higher animals. We build a mathematical framework to calculate how perturbations in maintenance and synthesis costs translate into contributions to the selection coefficient, a measure of relative fitness. This allows us to show that the hypothesis is an approximation to the actual baseline selection coefficient. Moreover we can directly derive the correct prefactor in its functional form, as well as analytical bounds on the accuracy of the hypothesis for any given realization of the model. We illustrate our general framework using a special case of the growth model, which we show provides a quantitative description of overall metabolic synthesis and maintenance expenditures in data collected from a wide array of unicellular organisms (both prokaryotes and eukaryotes). In all these cases we demonstrate that the hypothesis is an excellent approximation, allowing estimates of baseline selection coefficients to within 15\% of their actual values. Even in a broader biological parameter range, covering growth data from multicellular organisms, the hypothesis continues to work well, always within an order of magnitude of the correct result. Our work thus justifies its use as a versatile tool, setting the stage for its wider deployment.

\end{abstract}


\maketitle

 Discovering optimality principles in biological function has been a
 major goal of biophysics
 ~\cite{Dekel2005,dill2011physical,ten2016fundamental,hathcock2016noise,zechner2016molecular,fancher2017fundamental},
 but the competition between genetic drift and natural selection means
 that evolution is not purely an optimization
 process~\cite{Kimura1962,Ohta1973,Ohta1996}.  A necessary complement
 to elucidating optimality is clarifying under what circumstances
 selection is actually strong enough relative to drift in order to
 drive systems toward local optima in the fitness landscape.  In this
 work we focus on one key component of this problem: quantifying the
 selective pressure on the extra metabolic costs associated with a
 genetic variant.  We validate a long hypothesized
 relation~\cite{Orgel1980,Wagner2005,Lynch2015} between this pressure
 and the fractional change in the total resting metabolic expenditure
 of the organism.

The effectiveness of selection versus drift hinges on two
non-dimensional parameters~\cite{Gillespie2010}: i) the
{\it selection coefficient} $s$, a measure of the fitness of the
mutant versus the wild-type.  Mutants will have on average $1+s$
offspring relative to the wild-type per wild-type generation time;
ii) the {\it effective population} $N_e$ of the organism,
the size of an idealized, randomly mating population that exhibits the
same decrease in genetic diversity per generation due to drift as the
actual population (with size $N$).  For a deleterious mutant ($s<0$)
where $|s| \gg N_e^{-1}$, natural selection is dominant, with the
probability of the mutant fixing in the population exponentially
suppressed.  In contrast if $|s| \ll N_e^{-1}$, drift is dominant,
with the fixation probability being approximately the same as for a
neutral mutation~\cite{Kimura1962}.  Thus the magnitude of $N_e^{-1}$
determines the ``drift barrier''~\cite{Sung2012}, the critical minimum
scale of the selection coefficient for natural selection to play a
non-negligible role.

The long-term effective population size $N_e$ of an organism is
typically smaller than the instantaneous actual $N$, and can be
estimated empirically across a broad spectrum of life: it varies from
as high as $10^{9} - 10^{10}$ in many bacteria, to $10^{6} - 10^{8}$
in unicellular eukaryotes, down to $\sim 10^{6}$ in invertebrates and
$\sim 10^{4}$ in vertebrates~\cite{Lynch2015,Charlesworth2009}.  The
corresponding six orders of magnitude variation in the drift barrier
$N_e^{-1}$ has immense ramifications for how we understand selection
in prokaryotes versus eukaryotic organisms, particularly in the
context of genome complexity~\cite{Lynch2003,Lynch2005,Koonin2016}.
For example, consider a mutant with an extra genetic sequence relative
to the wild-type.  We can separate $s$ into two contributions, $s =
s_c + s_a$~\cite{Lynch2015}: $s_c$ is the baseline selection
  coefficient associated with the metabolic costs of having this
  sequence, i.e. the costs of replicating it during cell division,
  synthesizing any associated mRNA / proteins, as well as the
  maintenance costs associated with turnover of those components;
  $s_a$ is the correction due to any adaptive consequences of the
  sequence beyond its baseline metabolic costs.  For a prokaryote
with a low drift barrier $N_e^{-1}$, even the relatively low costs
associated with replication and transcription are often under
selective pressure~\cite{Wagner2005,Lynch2015}, unless $s_c < 0$ is
compensated for an $s_a >0$ of comparable or larger
magnitude~\cite{Sela2016}.  For the much greater costs of translation,
the impact on growth rates of unnecessary protein production is large
enough to be directly seen in experiments on
bacteria~\cite{Dekel2005,Scott2010}.  In contrast, for a eukaryote
with sufficiently high $N_e^{-1}$, the same $s_c$ might be effectively
invisible to selection, even if $s_a = 0$.  Thus even genetic
  material that initially provides no adaptive advantage can be
readily fixed in a population, making eukaryotes susceptible to
non-coding ``bloat'' in the genome.  But this also provides a rich
palette of genetic materials from which the complex variety of
eukaryotic regulatory mechanisms can subsequently
evolve~\cite{Taft2007,Lynch2015}.

Part of the explanatory power of this idea is the fact that the $s_c$
of a particular genetic variant should in principle be predictable
from underlying physical principles.  In fact, a very plausible
hypothesis is that $s_c \approx -\delta C_T/C_T$, where $C_T$ is the
total resting metabolic expenditure of an organism per generation
time, and $\delta C_T$ is the extra expenditure of the mutant versus
the wild-type.  This relation can be traced at least as far back as
the famous ``selfish DNA'' paper of Orgel and Crick~\cite{Orgel1980},
where it was mentioned in passing.  But its true usefulness was only
shown more recently, in the notable works of Wagner~\cite{Wagner2005}
on yeast and Lynch \& Marinov~\cite{Lynch2015} on a variety of
prokaryotes and unicellular eukaryotes.  By doing a detailed
biochemical accounting of energy expenditures, they used the relation
to derive values of $s_c$ that provided intuitive explanations of the
different selective pressures faced by different classes of organisms.
The relation provides a Rosetta stone, translating metabolic costs into evolutionary terms. And its full potential is
still being explored, most recently in describing the energetics of
viral infection~\cite{Mahmoudabadi2017}.

Despite its plausibility and long
pedigree, to our knowledge this relation has never been justified in complete generality from first principles.  We do so through a general
bioenergetic growth model, versions of which have been applied
across the spectrum of life~\cite{West2001,Hou2008,Kempes2012}, from
unicellular organisms to complex vertebrates.  We show that the relation is universal to an excellent approximation
across the entire biological parameter range.

{\it Growth model:} Let $\Pi(m(t))$ [unit: W] be the average power input into the resting metabolism of an organism (the metabolic expenditure after locomotion and other activities are accounted for~\cite{Hou2008}).  $\Pi(m(t))$ can be an
arbitrary function of the organism's current mass $m(t)$ [unit: g] at
time $t$.  This power is partitioned into maintenance of
existing biological mass (i.e. the turnover energy costs associated
with the constant replacement of cellular components lost to
degradation), and growth of new mass (i.e. synthesis of additional
components during cellular replication)~\cite{Pirt1965}.  Energy conservation implies
\begin{equation}\label{1}
  \Pi(m(t)) = B(m(t)) m(t) + E(m(t)) \frac{dm}{dt},
  \end{equation}
Here $B(m(t))$ [unit: W/g] is the maintenance cost per unit mass, and
$E(m(t))$ [unit: J/g] is the synthesis cost per unit
mass.  We allow both these quantities to be arbitrary
functions of $m(t)$.

Though we will derive our main result for the fully general model of
Eq.~\eqref{1}, we will also explore a special case: $\Pi(m(t)) = \Pi_0
m^{\alpha}(t)$, $B(m(t)) = B_m$, $E(m(t)) = E_m$, with scaling
exponent $\alpha$ and constants $\Pi_0$, $B_m$, and
$E_m$~\cite{Kempes2012}.  Allometric scaling of $\Pi(m(t))$ with
$\alpha = 3/4$ across many different species was first noted in the
work of Max Kleiber in the 1930s~\cite{Kleiber1932}, and with the
assumption of time-independent $B(m(t))$ and $E(m(t))$ leads to a
successful description of the growth curves of many higher
animals~\cite{West2001,Hou2008}.  However, recently there has been
evidence that $\alpha=3/4$ may not be universal~\cite{Delong2010,Ballesteros2018}.
Higher animals still exhibit $\alpha<1$ (with debate over
$\alpha = 2/3$ versus 3/4~\cite{White2003}), but unicellular
organisms have a broader range $\alpha \lesssim 2$.  Thus we will use
the model of Ref.~\cite{Kempes2012} with an arbitrary
species-dependent exponent $\alpha$.  While the resulting
  description is reasonable as a first approximation, particularly for
  unicellular organisms, one can easily imagine scenarios where the
  exponent and maintenance costs might vary between different developmental
  stages~\cite{Glazier2005}.  For the case of maintenance in
  endothermic animals, which in our approach includes all
  non-growth-related expenditures, more energy per unit mass is
  allocated to heat production as the organism
  matures~\cite{Werner2018}, effectively increasing the cost of
  maintenance.  In the Supplementary Information (SI) Sec.~V we show
  how the generalized model works in this scenario, using experimental
  growth data from two endothermic bird species~\cite{Dietz1997}.
  Thus it is useful to initially consider the model in complete
  generality.

{\it Baseline selection coefficient for metabolic costs:} To derive an
expression for $s_c$ for the growth model of Eq.~\eqref{1}, we first
focus on the generation time $t_r$, since this will be affected by alterations in metabolic costs.  $t_r$ is the typical age of reproduction, defined explicitly for any population model in SI Sec.~I, where we relate it to the population birth rate $r$ through $r = \ln (R_b)/t_r$~\cite{May1976,Savage2004}. Here $R_b$ is
the mean number of offspring per individual.  Let $\epsilon = m_r/m_0$ be the ratio of the mass $m_r = m(t_r)$ at reproductive maturity to the birth mass $m_0 = m(0)$.  For example in the case of symmetric binary fission of a unicellular organism, $R_b \approx \epsilon \approx 2$ (see SI Sec.~III for a discussion of $\epsilon$ in more general models of cell size homeostasis).  Since $m(t)$ is a
monotonically increasing function of $t$ for any physically realistic
growth model, we can invert Eq.~\eqref{1} to write the infinitesimal
time interval $dt$ associated with an infinitesimal increase of mass
$dm$ as $dt = dm\, E(m)/G(m)$ where $G(m) \equiv \Pi(m)-B(m)m$ is the
amount of power channeled to growth, and we have switched variables
from $t$ to $m$.  Note that $G(m)$ must be positive over the $m$ range
to ensure that $dm/dt>0$.  Integrating $dt$ gives us an expression for
$t_r$,
\begin{equation}\label{2}
t_r = \int_{m_0}^{\epsilon m_0} dm \frac{E(m)}{G(m)}.
\end{equation}
If we are interested in finding $s_c$ for a genetic variation, we can focus on the additional metabolic costs due to that variation.  For the purposes of calculation, this means treating the mutation as if it does not alter biological function in any other respect, including the ability of the organism to assimilate energy for its resting metabolism through uptake of nutrients or foraging.  If the mutation actually had only metabolic cost effects, the full selection coefficient $s=s_c$.  However generically mutations can affect both metabolic costs and power input (and/or other adaptive aspects), so $s=s_c +s_a$, with a correction term $s_a$ due to the adaptive effects~\cite{Lynch2015}.  In the latter case $s_c$ can still be calculated as shown below (ignoring adaptive effects) and interpreted as the baseline
  contribution to selection due to metabolic costs. While we do not focus on $s_a$ here, our theory can be readily extended to consider adaptive contributions as well, as illustrated in SI Sec.~VII, including aspects like spare respiratory capacity.  This broader formalism is summarized in Fig.~S3.
  
Proceeding with the $s_c$ derivation, the products of the genetic variation (i.e. extra mRNA transcripts or translated
proteins) may alter the mass of the mutant, which we denote by
$\tilde{m}(t)$.  The left-hand side of Eq.~\eqref{1} remains
$\Pi(m(t))$, where $m(t)$ is now the {\it unperturbed} mass of the
organism (the mass of all the pre-variation biological materials).
The power input $\Pi(m(t))$ depends on $m(t)$ rather than $\tilde
m(t)$ since only $m(t)$ contributes to the processes that allow the
organism to process nutrients, in accordance with the assumption
  that power input is unaltered in order to calculate $s_c$.  It is
also convenient to express our dynamics in terms of $m(t)$ rather than
$\tilde m(t)$, since the condition defining reproductive time $t_r$
remains unchanged, $m(t_r) = \epsilon m_0$, or in other words when the
unperturbed mass reaches $\epsilon$ times the initial unperturbed mass
$m_0$.  Thus Eq.~\eqref{1} for the mutant takes the form $\Pi(m(t)) =
\tilde B(m(t)) + \tilde E(m(t)) dm(t)/dt$, where $\tilde B(m(t)) =
B(m(t)) + \delta B$ and $\tilde E(m(t)) = E(m(t)) + \delta E$ are the
mutant maintenance and synthesis costs.  For simplicity, we
assume the perturbations $\delta B$ and $\delta E$ are independent of
$m(t)$, though this assumption can be relaxed.  In SI
  Sec.~IV, we show a sample calculation of $\delta B$ and $\delta E$
for mutations in ${\it E. coli}$ and fission yeast involving short
extra genetic sequences transcribed into non-coding RNA.  This
provides a concrete illustration of the framework we now develop.

Changes in the metabolic terms will perturb the generation
time, $\tilde t_r = t_r + \delta t_r$, and consequently the
birth rate $\tilde r = r+ \delta r$.  The corresponding
baseline selection coefficient $s_c$ can be exactly related to
$\tilde{s}_c \equiv -\delta t_r/t_r$, the fractional change in $t_r$,
through $s_c = R_b^{\tilde{s}_c/(1-\tilde{s}_c)}-1$ (see SI
Sec.~I).  This relation can be approximated as $s_c \approx
\ln(R_b) \tilde{s}_c$ when $|\tilde{s}_c| \ll 1$, the regime of
interest when making comparisons to drift barriers $N_e^{-1} \ll 1$.
In this regime $\tilde{s}_c \approx \delta r/r$, the fractional change
in birth rate.  While we focus here on the the simplest
  case of exponential population growth, where $\tilde{s}_c$ is
  time-independent, we generalize our approach to
  density-dependent growth models, where $\tilde{s}_c$ varies between
  generations, in SI Sec.~VI. $\tilde{s}_c$ can be written in a way
that directly highlights the contributions of $\delta E$ and $\delta
B$ to $\tilde{s}_c$.  To facilitate this, let us define the
  average of any function $F(m(t))$ over a single generation time
  $t_r$ as $\langle F \rangle \equiv t_r^{-1} \int_0^{t_r}
  dt\,F(m(t))$.  Changing variables from $t$ to $m$, like we did above
  in deriving Eq.~\eqref{2}, we can write this equivalently as
  $\langle F \rangle = \int_{m_0}^{\epsilon m_0} dm\,F(m) p(m)$, where
  $p(m) \equiv t_r^{-1} dt/dm = t_r^{-1} E(m)/G(m)$.  The value
  $p(m)dm$ is just the fraction of the generation time that the
  organism spends growing from mass $m$ to mass $m+dm$.  Expanding
Eq.~\eqref{2} for $t_r$ to first order in the perturbations $\delta E$
and $\delta B$, the coefficient $\tilde{s}_c = -\delta t_r/t_r =
-\sigma_E \delta E/\langle E \rangle -\sigma_B \delta B/\langle B
\rangle$, with positive dimensionless prefactors
\begin{equation}\label{3}
  \sigma_E \equiv \langle E \rangle \langle E^{-1} \rangle, \quad \sigma_B \equiv \langle B \rangle \langle \Theta^{-1} \rangle.
\end{equation}
Here $\Theta(m) \equiv G(m)/m$, and $F^{-1}(m) \equiv 1/F(m)$ for any $F$.  The
magnitude of $\sigma_B$ versus $\sigma_E$ describes how much
fractional increases in maintenance costs matter for selection
relative to fractional increases in synthesis costs.  We see that both prefactors are products of time averages of functions related to metabolism.  See SI Sec.~II for a detailed derivation of Eq.~\eqref{3}, and also Eq.~\eqref{5} below.

{\it Relating the baseline selection coefficient to the fractional
  change in total resting metabolic costs:} The final step in our
theoretical framework is to connect the above considerations to the
total resting metabolic expenditure $C_T$ of the organism per
generation time $t_r$, given by $C_T = \zeta \int_0^{t_r} dt\,
\Pi(m(t)) = \zeta t_r \langle \Pi \rangle$.  To compare with
the experimental data of Ref.~\cite{Lynch2015}, compiled in terms of
phosphate bonds hydrolyzed [P], we add the prefactor $\zeta$ which
converts from units of J to P.  Assuming an ATP hydrolysis energy of
50 kJ/mol under typical cellular conditions, we set $\zeta = 1.2\times
10^{19}$ P/J.  The genetic variation discussed above perturbs
  the total cost, $\tilde{C}_T = C_T + \delta C_T$, and the fractional
change $\delta C_T / C_T$ can be expressed in a form analogous to
$\tilde{s}_c$, namely $\delta C_T/C_T = \sigma^\prime_E \delta
E/\langle E \rangle +\sigma^\prime_B \delta B/\langle B \rangle$, with
\begin{equation}\label{5}
    \sigma^\prime_E \equiv \langle E \rangle \langle \Pi \rangle^{-1}\langle \Pi E^{-1} \rangle, \quad \sigma^\prime_B \equiv \langle B \rangle \langle \Pi \rangle^{-1} \langle \Pi \Theta^{-1} \rangle,
  \end{equation}
where again the prefactors are expressed in terms of time averages over metabolic functions. The connection between $s_c$ and $\delta C_T/C_T$ can
be constructed by comparing Eq.~\eqref{3} with
Eq.~\eqref{5}.  We see that $\tilde{s}_c = -\delta C_T/C_T$
for all possible perturbations $\delta E$ and $\delta B$ only when
$\sigma_E = \sigma_E^\prime$ and $\sigma_B = \sigma^\prime_B$.  We derive strict bounds on the differences between the prefactors (SI Sec.~II), which show that the relation is exact when: i) 
$\Pi(m)$ is a constant independent of $m$; and/or ii) $E(m)$ and
$\Theta(m)$ are independent of $m$.  Outside these cases, the relation
$\tilde{s}_c \approx -\delta C_T/C_T$ is an approximation.  To see how
well it holds, it is instructive to investigate the allometric growth
model described earlier, where $\Pi(m(t)) = \Pi_0 m^\alpha(t)$,
$E(m(t)) = E_m$, $B(m(t)) = B_m$.

\begin{figure}[t]
  \centering \includegraphics[width=\columnwidth]{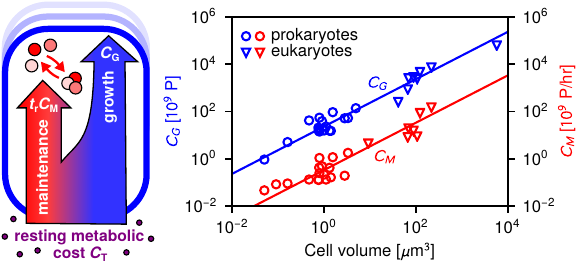}
  \caption{The growth $C_G$ (blue) and maintenance $C_M$ (red)
    contributions to an organism's total resting metabolic cost $C_T =
    C_G + t_r C_M$ per generation time $t_r$.  The symbols (circles
    $=$ prokaryotes, triangles $=$ unicellular eukaryotes) represent
    data tabulated in Ref.~\cite{Lynch2015}.  $C_G$ and $C_M$ have
    units of $10^9$ P (phosphate bonds hydrolyzed), and $10^9$ P/hr
    respectively. The lines represent best fits to the theoretical
    expressions for $C_G$ and $C_M$ from the allometric growth
    model.}\label{f1}
\end{figure}

{\it Testing the relation in an allometric growth model.}  We
  use model parameters based on the metabolic data of
Ref.~\cite{Lynch2015}, covering a variety of prokaryotes and
unicellular eukaryotes.  This data consisted of two quantities, $C_G$
and $C_M$, which reflect the growth and maintenance contributions to
$C_T$.  Using Eq.~\eqref{1} to decompose $\Pi(m(t))$, we can write
$C_T = C_G + t_r C_M$, where $C_G = \zeta \int_{m_0}^{\epsilon m_0}
dm\,E(m) = \zeta (\epsilon-1)m_0 E_m$ is the expenditure for growing
the organism, and $C_M = \zeta \langle B m \rangle = \zeta B_m \langle
m \rangle$ is the mean metabolic expenditure for maintenance per unit
time.  $C_G$ and $C_M$ scale linearly with cell
volume (SI Sec.~III), and best fits to the data, shown in Fig.~\ref{f1},
yield global interspecies averages: $E_m = 2,600$ J/g and $B_m
= 7 \times 10^{-3}$ W/g.  As discussed in the SI, these values are
remarkably consistent with earlier, independent estimates, for
unicellular and higher
organisms~\cite{Moses2008,Kempes2012,Maitra2015,Hou2008}.

\begin{figure}[t]
  \centering \includegraphics[width=\columnwidth]{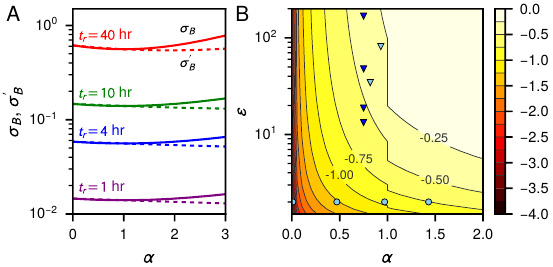}
  \caption{A: $\sigma_B$ (solid curves) from Eq.~\eqref{3} and
    $\sigma_B^\prime$ (dashed curves) from Eq.~\eqref{5} versus
    $\alpha$, for the allometric growth model with $E_m =
    2,600$ J/g, $B_m = 7 \times 10^{-3}$ W/g, and $\epsilon
    =2$.  At any given $\alpha$, the parameter $\Pi_0$ for each pair
    of curves (different colors) is chosen to correspond to particular
    reproductive times $t_r$, indicated in the labels. B: Contour
    diagram showing the logarithm of the maximum possible discrepancy
    $\log_{10}\,|1-\sigma_B^\prime/\sigma_B|$ for any allometric
    growth model parameters, as a function of $\alpha$ and $\epsilon$.
    To illustrate biological ranges $\alpha$ and $\epsilon$, the
    symbols correspond to data for various species (circles =
    unicellular, triangles = multicellular) drawn from the growth
    trajectories analyzed in Ref.~\cite{Kempes2012} (light blue) and
    Ref.~\cite{West2001} (dark blue).  SI Sec.~III shows a detailed
    species list.}\label{f2}
\end{figure}

Since $E(m(t)) = E_m$ is a constant in the allometric growth model,
$\sigma_E = 1$ from Eq.~\eqref{3}, and $\sigma_E =
  \sigma^\prime_E$ holds exactly from Eq.~\eqref{5}.  So the only
aspect of the approximation that needs to be tested is the similarity
between $\sigma_B$ and $\sigma_B^\prime$.  Fig.~\ref{f2}A shows
$\sigma_B$ versus $\sigma_B^\prime$ for the range $\alpha = 0 - 3$,
which includes the whole spectrum of biological
scaling~\cite{Delong2010} up to $\alpha =2$, plus some larger $\alpha$
for illustration.  For a given $\alpha$, the coefficient
$\Pi_0$ has been set to yield a certain division time $t_r =1 - 40$ hr, encompassing both the fast and slow extremes of typical
unicellular reproductive times.  In all cases $\sigma_B^\prime$ is in
excellent agreement with $\sigma_B$.  For the range $\alpha \le 2$ the
discrepancy is less than $15\%$, and it is in fact zero at the special
points $\alpha =0$, $1$.  Clearly the approximation begins to
break down at $\alpha \gg 1$, but it remains sound in the biologically
relevant regimes.  Note that $\sigma_B$ values for $t_r=1$ hr are
$\sim 0.01$, reflecting the minimal contribution of maintenance
relative to synthesis costs in determining the selection coefficient
for fast-dividing organisms.  This
limit is consistent with microbial metabolic flux
theory~\cite{Berkhout2013}, where maintenance is typically
neglected, so $\tilde{s}_c = -\delta C_T/C_T$ exactly (since only $\sigma_E =
\sigma_{E^\prime} =1$ matters).  As $t_r$ increases, so does $\sigma_B$ and hence the
  influence of maintenance costs, so by $t_r = 40$ hr, $\sigma_B$ is
  comparable to $\sigma_E$.

To make a more comprehensive analysis of the validity of the
$\tilde{s}_c \approx -\delta C_T/C_T$ relation, we do a computational
search for the worst case scenarios: for each value of $\alpha$ and
$\epsilon$, we can numerically determine the set of other growth model
parameters that gives the largest discrepancy
$\left|1-\sigma_B^\prime/\sigma_B\right|$.  Fig.~\ref{f2}B shows a
contour diagram of the results on a logarithmic scale,
$\log_{10}\,\left|1-\sigma_B^\prime/\sigma_B\right|$, as a function of
$\alpha$ and $\epsilon$.  Estimated values for $\alpha$ and $\epsilon$
from the growth trajectories of various species are plotted as symbols
to show the typical biological regimes.  While the maximum
discrepancies are smaller for the parameter ranges of unicellular
organisms (circles) compared to multicellular ones (triangles), in all
cases the discrepancy is less than $50\%$.  To observe a serious
error ($\sigma_B^\prime$ a different order of
magnitude than $\sigma_B$), one must go to the large $\alpha$, large
$\epsilon$ limit (top right of the diagram) which no longer
corresponds to biologically relevant growth trajectories.

{\it Validity of the relation in more complex growth scenarios:}
  Going beyond the simple allometric model, SI Sec.~V analyzes avian
  growth data, where the metabolic scaling exponent varies between
  developmental stages.  We find $\sigma_E = \sigma_E^\prime =1$ and
  the discrepancy $|1-\sigma_B^\prime/\sigma_B| \le 30\%$.  SI Sec.~VI
  considers density-dependent growth, illustrated by examples
  of bacteria competing for a limited resource in a chemostat and
  predators competing for prey.  Remarkably, when these systems approach a
  stationary state in total population and resource/prey
  quantity, we find $\sigma_E = \sigma_E^\prime =1$, $\sigma_B =
  \sigma_B^\prime = (B_m \ln R_b)/(E_m d \ln \epsilon)$, where $d$ is
  the dilution rate in the chemostat, or the predator death rate.  The
  simple expression for $\sigma_B$ allows straightforward estimation
  of the maintenance contribution to selection.  For the chemostat
  that contribution can be tuned experimentally through the dilution
  rate $d$.

We thus reach the conclusion that the baseline selection coefficient
for metabolic costs can be reliably approximated as $s_c \approx
-\ln(R_b) \delta C_T/C_T$.  As in the original
hypothesis~\cite{Orgel1980,Wagner2005,Lynch2015}, $-\delta C_T/C_T$ is
the dominant contribution to the scale of $s_c$, with corrections
provided by the logarithmic factor $\ln(R_b)$. Our derivation puts the
relation for $s_c$ on a solid footing, setting the stage for its wider
deployment.  It deserves a far greater scope of applications beyond
the pioneering studies of
Refs.~\cite{Wagner2005,Lynch2015,Mahmoudabadi2017}.  Knowledge of
$s_c$ can also be used to deduce the adaptive contribution $s_a =
s-s_c$ of a mutation, which has its own complex connection to
metabolism~\cite{Price2016} (see also SI Sec.~VII). The latter requires measurement of the
overall selection coefficient $s$, for example from competition/growth
assays, and the calculation of $s_c$ from the relation, assuming the
underlying energy expenditures are well characterized.  The $s_c$
relation underscores the key role of thermodynamic costs in shaping
the interplay between natural selection and genetic drift.  Indeed, it
gives impetus to a major goal for future research: a comprehensive
account of those costs for every aspect of biological function, and
how they vary between species, what one might call the
``thermodynome''.  Relative to its more mature omics brethren---the
genome, proteome, transcriptome, and so on---the thermodynome is still
in its infancy, but fully understanding the course of evolutionary
history will be impossible without it.

\let\oldaddcontentsline\addcontentsline
\renewcommand{\addcontentsline}[3]{}

\begin{acknowledgments}
The authors thank useful correspondence with M. Lynch, and feedback
from B. Kuznets-Speck, C. Weisenberger, and R. Snyder.  E.I. acknowledges support
from Institut Curie.
  \end{acknowledgments}

\nocite{*}
\let\addcontentsline\oldaddcontentsline

\appendix

\renewcommand{\theequation}{S\arabic{equation}}
\renewcommand{\thefigure}{S\arabic{figure}}
\renewcommand{\thetable}{S\arabic{table}}
\setcounter{equation}{0}
\setcounter{figure}{0}
\setcounter{section}{0}

\newpage

\begin{center}
  {\Large\bf Supplementary Information}
\end{center}

\tableofcontents

\section{I. Derivation of the relation between $s_c$ and $\tilde{s}_c$}\label{sc}

In the main text we posited an approximate relation $s_c \approx
\ln(R_b) \tilde{s}_c$ between the baseline selection coefficient $s_c$
and the fractional change in growth rate $\tilde{s}_c$ due to a
genetic variation.  Here we derive an exact relation between the two
quantities, generalizing the approach used in Ref.~\cite{Chevin2011}
for the specific case of binary fission. We then show how the
approximation used in the main text arises in the limit $|\tilde{s}_c|
\ll 1$.

Consider a group of wild-type organisms with population $N_w(t)$ as a
function of time.  Over each generation time $t_r$ the population grows by a factor of $R_0$, the net reproductive rate, so that after $n$ generations,
\begin{equation}\label{gr1}
    N_w(n t_r) = (R_0) ^{n} N_w(0).
\end{equation}
Here we focus on the simple case of exponential growth where both $R_0$ and $t_r$ are time-independent.  Below in Sec.~VI we consider more complex scenarios in systems with density-dependent population growth.  For a growing population $R_0>1$.  Most generally $R_0$ and $t_r$ are defined through $R_0 = \int_{0}^\infty dt\, p(t) f(t)$ and $t_r = R_0^{-1} \int_{0}^\infty dt\, t p(t) f(t)$, where $p(t)$ is the probability of the organism to survive from birth to age $t$, and $f(t)$ is the average fecundity at age $t$~\cite{May1976,Savage2004}.  A common simplification is to assume $f(t)$ is sharply peaked around age $t_r$, $f(t) \approx R_b \delta(t-t_r)$ with mean number of offspring $R_b$, so that $R_0 \approx R_b p(t_r)$~\cite{Savage2004, Kempes2012}.  One example is binary fission in bacteria, where $R_b =2$, and $p(t_r) \approx 1$ if one neglects cell deaths (and there is no cell removal), so $R_0 \approx R_b = 2$. In the continuum time approximation to population growth, $n = t/t_r$ and Eq.~\eqref{gr1} becomes
\begin{equation}\label{gr2}
\begin{split}
    N_w(t) = R_0^{t/t_r} N_w(0) &= e^{(t/t_r) \ln R_0} N_w(0)\\
    &\approx e^{(t/t_r) \ln (R_b p(t_r))}N_w(0)\\ &\equiv e^{(r-d)t}N_w(0)
    \end{split}
\end{equation}
describing Malthusian growth with net rate $r-d$, the difference between the birth rate $r = \ln(R_b)/t_r$ and death rate $d = -\ln(p(t_r))/t_r$.  

Now consider the growth of a mutant population $N_m(t)$.  Under the baseline assumption on which we focus in the main text (neglecting any adaptive effects), the mutant has the same mean number of offspring $R_b$ and death rate $d$, but metabolic cost differences affecting growth perturb the mean generation time to $\tilde{t}_r = t_r + \delta t_r$.  This leads to a modified birth rate $\tilde{r} = r+ \delta r = \ln(R_b)/(t_r+\delta t_r)$, and a growth equation
\begin{equation}\label{gr3}
N_m(t) = e^{(\tilde{r} - d)t} N_m(0).
\end{equation}
From Eqs.~\eqref{gr2}-\eqref{gr3} and the definitions of $r$ and $\tilde{r}$, the ratio of the mutant to wild type populations is given by
\begin{equation}\label{s1}
\frac{N_m (t)}{N_w (t)}= \frac{N_m (0)}{N_w (0)}e^{t \delta r} = \frac{N_m (0)}{N_w (0)}R_b ^{t\left (\frac{1}{t_{r}+\delta t_r}-\frac{1}{t_{r}} \right)}.
\end{equation}
On the other hand, after $n$
wild-type generations ($t = n t_r$) the ratio of the two populations
is related to the selection coefficient (as conventionally defined in
population genetics) through
\begin{equation}\label{s2}
\frac{N_m (n t_r)}{N_w (n t_r)}=\frac{N_m (0)}{N_w (0)}(1+s_c)^{n}.
\end{equation}
Plugging $t = n t_r$ into Eq.~\eqref{s1} and comparing to
Eq.~\eqref{s2}, we see that
\begin{equation}\label{s3}
  s_c = R_b^{-\frac{\delta t_r}{t_r + \delta t_r}} -1.
\end{equation}
If we define $\tilde{s}_c \equiv -\delta t_r / t_r$, then
Eq.~\eqref{s3} can be rewritten as 
\begin{equation}\label{s4}
  s_c = R_b^{\frac{\tilde{s}_c}{1-\tilde{s}_c}} -1.
\end{equation}
Note that in the case where $|\tilde{s}_c| \ll 1$, or $|\delta t_r| \ll
t_r$, we can also write $\tilde{s}_c \approx \delta r / r$, and so
interpret $\tilde{s}_c$ as the fractional change in birth rate.  In
this same limit we can expand Eq.~\eqref{s4} for small $\tilde{s}_c$,
\begin{equation}\label{s5}
s_c=\ln(R_b) \tilde{s}_c +\frac{1}{2} \ln( R_b)\left(2+\ln( R_b) \right)\tilde{s}_c ^2+ \cdots.
\end{equation}
Keeping only the leading order term, linear in $\tilde{s}_c$, yields
the approximation $s_c \approx \ln(R_b) \tilde{s}_c$.

\section{II. Derivation of main text Eqs.~(3)-(4) and related bounds}\label{der}

To derive Eq.~(3) in the main text, we start with the equation for $t_r$ [main text Eq.~(2)]:
\begin{equation}\label{dd1}
t_r = \int_{m_0}^{\epsilon m_0} dm \frac{E(m)}{G(m)} = \int_{m_0}^{\epsilon m_0} dm \frac{E(m)}{\Pi(m) - B(m) m},
\end{equation}
with $G(m) \equiv \Pi(m) - B(m) m$.  Under the perturbations $E(m) \to E(m) + \delta E$ and $B(m) \to B(m) +\delta B$, the generation time is altered to
\begin{equation}\label{dd2}
\begin{split}
t_r + \delta t_r &= \int_{m_0}^{\epsilon m_0} dm \frac{E(m)+\delta E}{\Pi(m) - (B(m)+\delta B) m}\\
&\approx t_r + \delta E \int_{m_0}^{\epsilon m_0} dm \frac{1}{\Pi(m) - B(m) m} + \delta B \int_{m_0}^{\epsilon m_0} dm \frac{m E(m)}{(\Pi(m) - B(m) m)^2},
\end{split}
\end{equation}
where in the second line we have Taylor expanded the integrand to first order in $\delta E$ and $\delta B$.  We can thus write $\tilde{s}_c \equiv -\delta t_r/t_r$ as
\begin{equation}\label{dd3}
    \tilde{s}_c = -\frac{\delta t_r}{t_r} = -\frac{\delta E}{t_r} \int_{m_0}^{\epsilon m_0} dm \frac{1}{G(m)} - \frac{\delta B}{t_r} \int_{m_0}^{\epsilon m_0} dm \frac{m E(m)}{G^2(m)}.
\end{equation}
Using the definitions $p(m) \equiv t_r^{-1} E(m)/G(m)$, $\langle F \rangle \equiv \int_{m_0}^{\epsilon m_0} dm\,F(m)p(m)$, and $\Theta(m) \equiv G(m)/m$, we can rewrite Eq.~\eqref{dd3} as
\begin{equation}\label{dd4}
    \tilde{s}_c = -\langle E^{-1} \rangle \delta E - \langle \Theta^{-1} \rangle \delta B,
\end{equation}
where $F^{-1}(m) \equiv 1/F(m)$ for any function $F$.  Finally we define the prefactors $\sigma_E$ and $\sigma_B$ through the relation
\begin{equation}\label{dd5}
    \tilde{s}_c = -\sigma_E \frac{\delta E}{\langle E \rangle} -\sigma_B \frac{\delta B}{\langle B \rangle}.
\end{equation}
Comparing Eq.~\eqref{dd4} to \eqref{dd5} we find the result of main text Eq.~(3):
\begin{equation}\label{dd6}
  \sigma_E = \langle E \rangle \langle E^{-1} \rangle, \quad \sigma_B = \langle B \rangle \langle \Theta^{-1} \rangle.
\end{equation}

The derivation of main text Eq.~(4) proceeds analogously, starting with the total resting metabolic expenditure per generation,
\begin{equation}\label{dd7}
C_T = \zeta \int_0^{t_r} dt\,\Pi(m(t)) = \zeta \int_{m_0}^{\epsilon m_0} dm\,\frac{dt}{dm} \Pi(m) = \zeta \int_{m_0}^{\epsilon m_0} dm\, \frac{E(m) \Pi(m)}{\Pi(m) - B(m) m},
\end{equation}
where we have changed variables in the integral from $t$ to $m$ and used the fact that $dt/dm = E(m)/G(m)$.  Under the perturbations $E(m) \to E(m) + \delta E$ and $B(m) \to B(m) +\delta B$, the expenditure is altered to
\begin{equation}\label{dd8}
\begin{split}
C_T + \delta C_T &= \zeta \int_{m_0}^{\epsilon m_0} dm\, \frac{(E(m)+\delta E) \Pi(m)}{\Pi(m) - (B(m)+\delta B) m}\\
 &= C_T + \zeta \delta E \int_{m_0}^{\epsilon m_0} dm\, \frac{\Pi(m)}{\Pi(m) - B(m) m}+ \zeta \delta B \int_{m_0}^{\epsilon m_0} dm\, \frac{E(m)\Pi(m)m}{(\Pi(m) - B(m) m)^2}\\
 &= C_T + \zeta t_r \langle \Pi E^{-1}\rangle \delta E + \zeta t_r \langle \Pi \Theta^{-1}\rangle\delta B .
\end{split}
\end{equation}
Using the fact that Eq.~\eqref{dd7} can also be written as $C_T = \zeta t_r \langle \Pi \rangle$, we can use Eq.~\eqref{dd8} to express the ratio $\delta C_T / C_T$ as
\begin{equation}\label{dd9}
    \frac{\delta C_T}{C_T} = \frac{\langle \Pi E^{-1}\rangle}{\langle \Pi \rangle} \delta E + \frac{\langle \Pi \Theta^{-1}\rangle}{\langle \Pi \rangle}\delta B.
\end{equation}
Comparing this to the expression defining $\sigma^\prime_E$ and $\sigma^\prime_B$,
\begin{equation}\label{dd10}
    \frac{\delta C_T}{C_T} = \sigma_E^\prime \frac{\delta E}{\langle E \rangle} +\sigma_B^\prime \frac{\delta B}{\langle B \rangle},
\end{equation}
we find the result of main text Eq.~(4):
\begin{equation}\label{dd11}
    \sigma^\prime_E \equiv \langle E \rangle \langle \Pi \rangle^{-1}\langle \Pi E^{-1} \rangle, \quad \sigma^\prime_B \equiv \langle B \rangle \langle \Pi \rangle^{-1} \langle \Pi \Theta^{-1} \rangle.
  \end{equation}
The degree to which $\tilde{s}_c$ can be approximated as $\delta C_T/C_T$ depends on the similarity of the prefactor $\sigma_E$ to $\sigma_E^\prime$, and $\sigma_B$ to $\sigma_B^\prime$.  Their relative differences can be written as:
\begin{equation}\label{bound}
  \begin{split}
    \left|1 - \frac{\sigma_E^\prime}{\sigma_E}\right| &= \left|1-\frac{\langle \Pi E^{-1}\rangle}{\langle \Pi \rangle \langle E^{-1} \rangle}\right| \le \kappa(\Pi) \kappa(E^{-1}),\\
    \left|1 - \frac{\sigma_B^\prime}{\sigma_B}\right| &= \left|1-\frac{\langle \Pi \Theta^{-1}\rangle}{\langle \Pi \rangle \langle \Theta^{-1} \rangle}\right| \le \kappa(\Pi) \kappa(\Theta^{-1}).
    \end{split}
  \end{equation}
where $\kappa(F) \equiv \sqrt{\langle F^2 \rangle - \langle F
  \rangle^2}/\langle F \rangle$ and we have used the Cauchy-Schwarz
inequality.  These bounds imply two
cases when $\tilde{s}_c$ is exactly equal to $\delta C_T /C_T$: i) $\kappa(\Pi)=0$, which means $\Pi(m)$ is a constant independent of $m$; ii) $\kappa(\Pi) >0$ and
$\kappa(E^{-1}) = \kappa(\Theta^{-1}) = 0$, which means $E(m)$ and
$\Theta(m)$ are independent of $m$.

\section{III. Fitting of allometric growth model to experimental data}\label{fit}

As discussed in the main text, we can decompose $C_T$ into two
components, $C_T = C_G + t_r C_M$, where $C_G = \zeta
\int_{m_0}^{\epsilon m_0} dm\,E(m)$ is the expenditure for growing the
organism, and $C_M = \zeta \langle B m \rangle$ is the mean metabolic
expenditure for maintenance per unit time.  For the allometric growth
model, these contributions to $C_T$ simplify to $C_G = \zeta
(\epsilon-1)m_0 E_m$ and $C_M = \zeta B_m \langle m \rangle$.
Ref.~\cite{Lynch2015} noted that $C_G$ and $C_M$ collected from
experimental data scaled nearly linearly with cell volume, with
allometric exponents of $0.97 \pm 0.04$ and $0.88 \pm 0.07$
respectively.  In fact, the simplest version of the allometric model
predicts exactly linear scaling, using the following assumptions.
Since the data tabulated in Ref.~\cite{Lynch2015} covers prokaryotes
and unicellular eukaryotes, we take $\epsilon =2$. Since the mass of
the organism varies between $m_0$ and $2m_0$ over time $t_r$, we
approximate $\langle m \rangle \approx (3/2) m_0$.  Note that setting $\epsilon =2$ assumes symmetric binary fission, though unicellular organisms can also exhibit asymmetric fission where $\epsilon \ne 2$~\cite{Marantan2016}.  However since the variation in $\epsilon$ is typically within a factor of two of the symmetric case, any errors introduced by this assumption, and the approximation for $\langle m \rangle$, will not change the order of
magnitude of the estimated model parameters.  For simplicity we are also ignoring variation in birth sizes $m_0$ over the population, which in the unicellular case is closely related to the possible mechanisms of cell size homeostasis, so-called sizer, adder, or timer behaviors~\cite{Facchetti2017,Sauls2016}.  In the current context, once a steady state size distribution is reached, we can define $\epsilon$ as the ratio of mean size at division to the mean birth size $m_0$, which will have different interpretations depending on the mechanism.  In the sizer picture $\epsilon = m_\text{r}/m_0$, where $m_\text{r}$ is the target cell mass at which division occurs.  In the adder picture, where cells divide after adding a mass $\Delta m$, we have $\epsilon = 1+\Delta m / m_0$.  In the timer mechanism, where division occurs after a specified time, additional conditions are needed to achieve a steady state size distribution, for example linear growth (in which case timer and adder are equivalent) or a mixed model coupling sizer and timer behavior~\cite{Grilli2017}.  In all cases, capturing the full effects of these mechanisms would require modeling the variation in birth sizes and their evolution over time, which we leave for a future work.

We relate the
experimentally observed cell volume $V$ to the mean cell mass $\langle
m \rangle$ by assuming a typical cell is 2/3 water (density
$\rho_\text{wat} = 10^{-12}\:\text{g}/\mu\text{m}^3$) and 1/3 dry
biomass (density $\rho_\text{dry} \approx 1.3\times
10^{-12}\:\text{g}/\mu\text{m}^3$)~\cite{bionum}.  Hence $\langle m
\rangle = (2\rho_\text{wat} + \rho_\text{dry})V/3 \equiv
\rho_\text{cell} V$.  We thus find:
\begin{equation}\label{g1}
C_G = (2/3)\zeta E_m \rho_\text{cell} V, \qquad C_M = \zeta B_m \rho_\text{cell} V.
\end{equation}
For each expression we have only one unknown parameter, $E_m$ and $B_m$
respectively.  Best fits to the Ref.~\cite{Lynch2015} data, shown in
main text Fig.~1B, yield global interspecies averages of the parameters,
$E_m = 2,600$ J/g and $B_m = 7 \times 10^{-3}$ W/g.

The fitted values are consistent with earlier approaches, once water
content is accounted for (i.e. to get $E_m$ per dry biomass, multiply
the value by $\approx 3$, so $E_m^\text{dry} = 7,800$ J/g).  The
synthesis cost $E_m$ has a very narrow range across many species, with
$E_m = 1,100-1,800$ J/g in bird and fish embryos, and $4,000-7,500$
J/g for mammal embryos and juvenile fish, birds, and
mammals~\cite{Moses2008}.  This energy scale seems to persist down to
the prokaryotic level, with $E_m^{\text dry} = 3,345$ J/g estimated
for {\it E. coli}~\cite{Kempes2012}.  $E_m^\text{dry}$ also appears in
a different guise as the inverse of the ``energy efficiency''
$\varepsilon$ of {\it E. coli} growth in the model of
Ref.~\cite{Maitra2015}; converting the optimal observed $\varepsilon
\approx 15$ dry g/(mol ATP) yields $E_m^{\text dry} = \zeta /
\varepsilon = 3,333$ J/g, consistent with the other estimates cited
above, as well as our fitted value.  The ratio $B_m/E_m$ was estimated
for various species in Ref.~\cite{Kempes2012}, and found to vary in
the range $10^{-6} - 10^{-5}$ s$^{-1}$ from prokaryotes to unicellular
eukaryotes, entirely consistent with our fitted value of $B_m/E_m =
3\times 10^{-6}$ s$^{-1}$.  The scale shifts for larger, multicellular
species, but not dramatically.  For example for a subset of mammals
with scaling $\alpha = 3/4$, adult mass sizes $m_a = 10 - 6.5\times
10^5$ g, and typical values of $\Pi_0 \approx 0.022$ W/g$^{3/4}$, $E_m
\approx 7000$ J/g~\cite{Hou2008}, we get a range of $B_m/E_m = 10^{-7}
- 10^{-6}$ s$^{-1}$.  We thus have confidence that the growth model
provides a description of the metabolic expenditures (in terms of
growth and maintenance contributions) that is consistent both with
the empirical data of Ref.~\cite{Lynch2015} and parameter expectations
based on a variety of earlier approaches.

For the symbols in the contour diagram of main text figure Fig.~2B, we
used parameters extracted from growth trajectories analyzed in
Ref.~\cite{Kempes2012} (light blue) and Ref.~\cite{West2001} (dark
blue).  Circles (left to right) are unicellular organisms ($\epsilon
=2$): {\it T. weissflogii}, {\it L. borealis}, {\it B. subtilis}, {\it
  E. coli}.  Triangles (top to bottom) are multicellular organisms:
guinea pig, {\it C. pacificus}, hen, {\it Pseudocalanus sp.}, guppy,
cow.  For the multicellular case the plotted values of $\epsilon$
correspond to asymptotic adult mass in units of $m_0$.  This is an
upper bound on $\epsilon$, though the actual $\epsilon$ should
typically be comparable~\cite{West2001,Ricklefs2010}.

\section{IV. Sample calculation of the baseline selection coefficient: short, non-coding RNA in {\it E. coli} and fission yeast}\label{example}

To illustrate a calculation of baseline selection coefficients in the
framework developed in the main text, let us consider a specific
biological example: a mutant with a short ($< 200$ bp) sequence in the
genome that is transcribed into non-coding RNA, and which is not
present in the wild-type.  We will focus on two organisms, the
prokaryote {\it E. coli} and the unicellular eukaryote
{\it S. pombe} (fission yeast).  To date we know that at least some
subset of non-coding RNA transcripts have functional roles in these
organisms~\cite{Raghavan2011,Leong2014}.  The evolution of such
regulatory sequences will be shaped both by the selective advantage
$s_a$ of having the sequence in the genome, and the baseline
disadvantage $s_c$ from the extra energetic costs of copying and
transcription.

Before calculating $s_c$, we first establish the validity of the
growth model for these organisms.  The model parameters fitted for the
data from prokaryotes and unicellular eukaryotes in Fig. 1B of the
main text are $E_m = 2,600$ J/g and $B_m = 7 \times 10^{-3}$ W/g.  The
corresponding growth and maintenance contributions to the total
resting metabolic cost per generation, $C_G$ and $C_M$, are given by
Eq.~\eqref{g1}.  Using $\zeta = 1.2\times 10^{19}$ P/J (recall that P
corresponds to ATP or ATP equivalents hydrolyzed), $\rho_\text{cell} =
1.1 \times 10^{-12}$ g/$\mu$m$^3$, and typical cell volumes
$V_\text{\it E.coli} = 1$ $\mu$m$^3$~\cite{bionum}, $V_\text{\it
  S.pombe} = 106$ $\mu$m$^3$~\cite{Chang2017}, we find: $C_G^{\it
  E.coli} = 2.30 \times 10^{10}$ P, $C_M^{\it E.coli} = 3.34 \times
10^8$ P/hr, $C_G^{\it S.pombe} = 2.43 \times 10^{12}$ P, $C_M^{\it
  S.pombe} = 3.54 \times 10^{10}$ P/hr.  These agree well in magnitude
with the literature estimates compiled in the SI of
Ref.~\cite{Lynch2015} (all normalized to $20^\circ$C): $C_G^{\it
  E.coli} = 1.57 \times 10^{10}$ P, $C_M^{\it E.coli} = 2.13 \times
10^8$ P/hr, $C_G^{\it S.pombe} = 2.35 \times 10^{12}$ P, $C_M^{\it
  S.pombe} = 8.7 \times 10^{9}$ P/hr.  Thus the globally fitted $E_m$
and $B_m$ values are physically reasonable for both organisms.

The extra sequence in the mutant leads to perturbations in both
synthesis cost per unit mass, $\delta E$, and maintenance cost per
unit mass, $\delta B$.  To calculate the first, we use the following
estimates based on the analysis in Ref.~\cite{Lynch2015}: for a
sequence of length $L$, the total DNA-related synthesis cost is $d_\xi
L$, where the label $\xi =$ {\it E. coli} or {\it S. pombe}.  Here the prefactor
$d_{\it E.coli} \approx 101$ P and $d_{\it S.pombe} \approx 263$ P.
If the steady-state average number of corresponding mRNA trascripts in
the cell is $N_r$, the additional ribonucleotide synthesis costs are
$\approx 46 N_r L$ in units of P.  Hence we have, per unit mass,
\begin{equation}\label{s6}
  \delta E \approx \frac{\zeta^{-1} L}{\rho_\text{cell} V_\xi} \left(d_\xi + 46 N_r \right),
\end{equation}
with the $\zeta^{-1}$ prefactor converting from P to J, so that
$\delta E$ has units of J/g.  The same analysis~\cite{Lynch2015}
yields the maintenance cost per unit time for replacing transcripts
after degradation, $\approx 2 N_r L \gamma_\xi$ in units of P/s, where
$\gamma_{\it E.coli} = 0.003$ s$^{-1}$ and $\gamma_{\it S.pombe} =
0.001$ s$^{-1}$ are the RNA degradation rates for the two organisms.
Per unit mass, the maintenance perturbation $\delta B$ is given by
\begin{equation}\label{s7}
  \delta B \approx \frac{2 \zeta^{-1} L N_r \gamma_\xi}{\rho_\text{cell} V_\xi},
\end{equation}
in units of W/g.

The final step is to calculate the prefactors $\sigma_E$ and
$\sigma_B$ from Eq.~(3) in the main text.  For this we need to choose
a particular growth model exponent $\alpha$, and we set $\alpha = 1$,
corresponding to the assumption of exponential cell mass growth.  In
this case $\sigma_E = 1$ for both organisms, while $\sigma^{\it
  E.coli}_B = 0.0070$, $\sigma^{\it S.pombe}_B = 0.060$.  The choice
of $\alpha$ has a minimal influence on the prefactors: $\sigma_E =1$
exactly for any model with a constant function $E(m) = E_m$.
Moreover, any $\alpha$ value in the biologically relevant range of $0 \le \alpha \le 2$ yields a $\sigma_B$ value within 5\% of the $\alpha = 1$ result
for each organism.

Putting everything together, we now can calulate all the components of
main text Eq.~(3) for $\tilde{s}_c$, namely $\sigma_E$, $\sigma_B$,
$\delta E$, $\delta B$, $\langle E \rangle = E_m$, and $\langle B
\rangle = B_m$.  Had we chosen instead to use the $\delta C_T/C_T$
approximation of main text Eq.~(4), the only discrepancy would have
been in the fact that $\sigma_B^\prime \ne \sigma_B$, since
$\sigma_E^\prime = \sigma_E = 1$.  However the discrepancy is small,
with $|1 - \sigma_B^\prime/\sigma_B| < 0.09$ for both organisms in the
range $0 \le \alpha \le 2$.

\begin{figure}[t]
  \includegraphics[width=\textwidth]{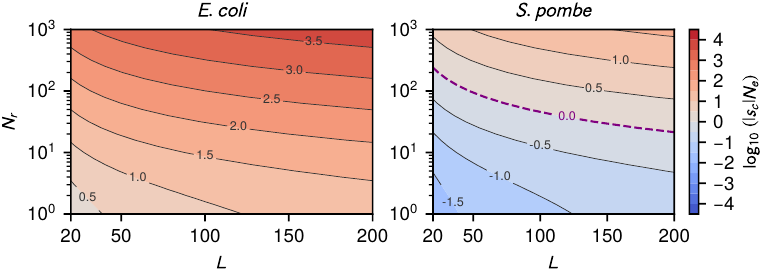}
  \caption{\linespread{1.0}\selectfont{} Contour diagrams of $\log_{10}(|s_c|N_e)$ as a function of sequence length $L$ and mean RNA transcript number $N_r$ per cell for {\it E. coli} (left) and {\it S. pombe} (right).  The dashed line in the diagram on the right corresponds to $|s_c| = N_e^{-1}$.}\label{fs1}
  \end{figure}

Fig.~\ref{fs1} shows contour diagrams of $\log_{10}(|s_c|N_e)$ as a
function of $L$ and $N_r$ for {\it E. coli} and {\it S. pombe}.  Here
$s_c = \ln(2) \tilde{s}_c$, assuming $R_b = 2$, and the effective
population sizes are $N_e^{\it E.coli} = 2.5\times
10^7$~\cite{Charlesworth2006}, $N_e^{\it S.pombe} = 1.2 \times
10^7$~\cite{Farlow2015}.  For ${\it E. coli}$, with its smaller
metabolic expenditures per generation relative to fission yeast, the
cost of the extra sequence is always significant: $|s_c| > N_e^{-1}$ for the
entire range of $L$ and $N_r$ considered, even for the smallest length
($L = 20$ bp) and a single transcript per cell on average, $N_r =1$.
Thus there will always be strong selective pressure to remove the
extra sequence, unless $s_c$ is compensated for by a comparable or
greater adaptive advantage $s_a$.  In contrast, for {\it S. pombe}
there is a regime of $L$ and $N_r$ where $|s_c| < N_e^{-1}$ (the region below
the dashed line).  Here the selective disadvantage of the extra
sequence is weaker than genetic drift, and such a genetic variant
could fix in the population at roughly the same rate as a neutral
mutation even if it conferred no selective advantage, $s_a = 0$.
While this makes fission yeast more tolerant of genomic ``bloat''
relative to {\it E. coli}, initially non-functional extra genetic material
could subsequently facilitate the development of novel regulatory
mechanisms.

\section{V. Generalized growth model with varying developmental scaling regimes}\label{bird}

In the last section of the main text, we explored the validity of the baseline selection coefficient relation in the simplest version of the growth model, with allometric power input $\Pi(m(t)) = \Pi_0 m^\alpha(t)$ and time-independent synthesis and maintenance factors $E(m(t)) = E_m$, $B(m(t)) = B_m$.  While this may be a reasonable approximation for various kinds of organisms~\cite{West2001,Hou2008,Kempes2012} (particularly those without major physiological / morphological changes during their development~\cite{Glazier2005}), there is evidence that in certain cases a single power law scaling with mass cannot accurately capture the resting metabolic rate (a detailed review is provided in Ref.~\cite{Glazier2005}).  This includes insects~\cite{Sears2012,Callier2012} and marine invertebrates~\cite{Glazier2018} that progress through several distinct developmental stages or instars, as well as endothermic birds when comparing juvenile and adult metabolism~\cite{Dietz1997}.  In these cases the resting metabolic input $\Pi(m(t))$ may scale approximately like a power-law in $m(t)$ during individual stages of development, but the exponent may vary from stage to stage.  Additionally, the rate of energy expenditure for maintenance, $B(m(t)) m(t)$ in our model, may also not have a simple time-independent prefactor $B(m(t)) = B_m$, since the cost of maintaining a unit of mass can also vary throughout development:  in juvenile endothermic birds, before internal heat production has reached its mature level and while there is no energy allocation toward reproductive functions, the maintenance costs are lower than in adult birds~\cite{Werner2018}.  There are also factors that influence metabolic scaling, for example the dimensionality of the environment in which the consumer organism encounters the resources on which it subsists~\cite{Pawar2012}.  Another example is a recent model where the partitioning of energy between metabolic processes and heat loss leads to a $\Pi(m)$ that is a linear combination of power law terms with different exponents (2/3 and 1)~\cite{Ballesteros2018}.  Thus it is important to have a general theory, one where $\Pi(m(t))$, $B(m(t))$ and even $E(m(t))$ could be arbitrary functions reflecting changes in the organism throughout development.  We note that the mathematical formalism developed in the main text, including Eqs.~(1-5), is indeed fully general, independent of any specific choice for these functions.

\begin{figure}[t]
  \includegraphics[width=\textwidth]{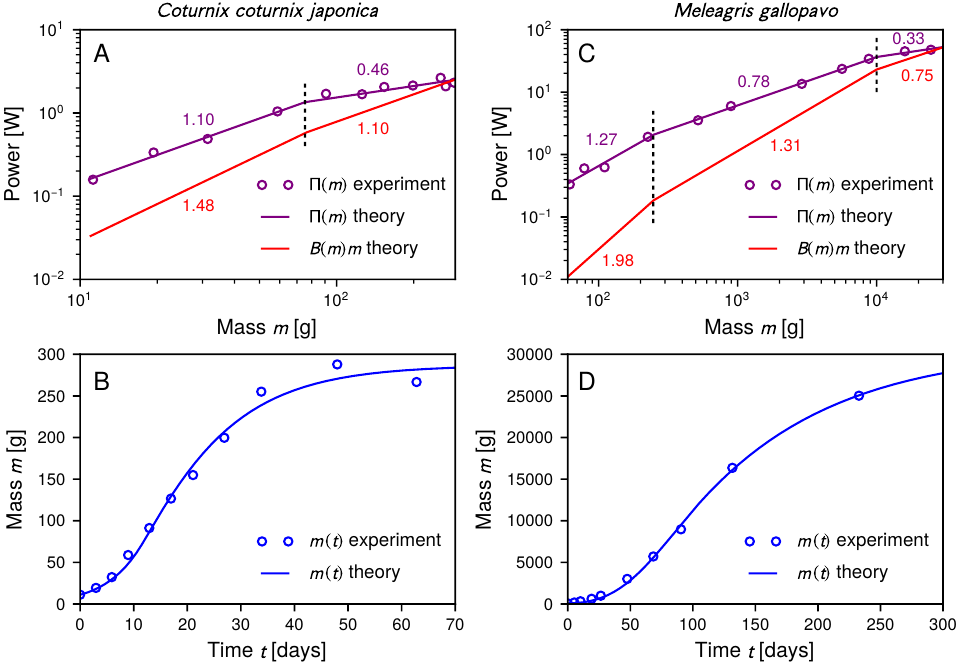}
  \caption{\linespread{1.0}\selectfont{} Fits of the generalized growth model to experimental results (taken from Ref.~\cite{Dietz1997}) for male members of two galliform species: Japanese quail ({\it Coturnix coturnix japonica}, left column); turkey ({\it Meleagris gallopavo}, right column).  A, C: Power input $\Pi(m)$ as a function of mass $m$ (purple), with symbols indicating experimental data and lines a piecewise continuous power-law fit to the data.  The power $B(m)m$ expended in maintenance is shown in red, and is calculated by fitting the growth model to the mass vs. time data in panel B.  Numbers next to the lines indicate the power-law exponents in each regime, while vertical dashed lines show the regime boundaries.  B, D: Mean mass $m$ versus time $t$, with symbols denoting experimental data and the curve the theoretical best-fit to the generalized growth model.}\label{fs2}
  \end{figure}

To illustrate how this formalism can be applied for more complex growth models with varying developmental regimes, we analyze empirical data~\cite{Dietz1997} from two endothermic, galliform bird species:  Japanese quail ({\it Coturnix coturnix japonica}) and turkey ({\it Meleagris gallopavo}).  Fig.~\ref{fs2} A,C shows the experimental resting metabolic rates $\Pi(m)$ (symbols) as a function of $m$.  Before applying our selection coefficient theory, we first establish the details of the growth model for each organism.  Following the analysis of Ref.~\cite{Dietz1997}, we fit $\Pi(m)$ using distinct power-law scaling exponents across different mass regimes (two regimes for quail, three for turkey).  The theory fits thus appear as piecewise continuous linear functions (purple lines) on the log-log graphs of panels A and C, with the exponents indicated above each regime.  The specific forms of the fitting functions are:
\begin{equation}
    \begin{split}
        \text{quail:} \qquad \Pi(m) &= \begin{cases} 0.0116\,m^{1.10} & m\le 75 \\
        0.185\, m^{0.46} & m > 75 \end{cases}\\
        \text{turkey:} \qquad \Pi(m) &= \begin{cases} 0.00191\, m^{1.27} & m\le 247 \\
        0.0288\, m^{0.78} & 247 < m \le 10^{4}\\
        1.70\, m^{0.33} & m>10^{4}
        \end{cases}
    \end{split}
\end{equation}
where the units of $m$ are g and of $\Pi(m)$ are W.  These forms for $\Pi(m)$ fit the experimental data closely, and exhibit a trend seen commonly in organisms with distinct metabolic scaling at different life stages:  the power-law exponent progressively decreases as the organism matures (see for example Type III and Type IV scaling behavior as classified by Glazier in Ref.~\cite{Glazier2005}).  And though the exponents vary between regimes, they all still fall in the typical biological range $\lesssim$ 2.

The resting metabolic power input determines the mass trajectory $m(t)$ of the organism through the energy conservation equation [Eq.~(1) in the main text]:
\begin{equation}\label{m1}
  \Pi(m(t)) = B(m(t)) m(t) + E(m(t)) \frac{dm}{dt}.
\end{equation}
Since we know the empirical $m(t)$ curves (symbols in Figs.~\ref{fs2} B,D), we can use these to find best-fit forms for $B(m(t))$ and $E(m(t))$.  For the latter we will assume the cost of synthesizing a new unit mass are constant throughout development, $E(m(t)) = E_m$.  This assumption is based on the fact that estimated $E_m$ values are broadly consistent among many different organisms at different life stages, as described in Sec.~III above, generally of the order of magnitude $E_m \sim {\cal O}(10^3)$ W/g.  For $B(m)$ we allow a more general fitting form, with the product $B(m) m$ (the maintenance power consumption) having piecewise continuous power-law scaling, with the same mass regimes as $\Pi(m)$.  Note that in our model $B(m) m$ subsumes all the parts of the resting metabolic power that are consumed in processes other than growth.  This is a broader definition of ``maintenance'' than in models which distinguish the uses of that power, for example separately keeping track of energy used for heat production and reproduction~\cite{Werner2018}.  However for our purposes it is sufficient to collectively track the total non-growth expenditures through $B(m)m$.  The time $t(m)$ to reach a mass $m$, starting from initial mass $m_0 = 11$ g (quail), $70$ g (turkey), is analogous to Eq.~(2) of the main text:
\begin{equation}\label{m2}
t(m) = \int_{m_0}^{m} dm \frac{E_m}{\Pi(m) - B(m) m}.
\end{equation}
By inverting this to get $m(t)$ and comparing to the empirical data for the mass trajectories, we can find theoretical best-fits for $E_m$ and the functional forms of $B(m)m$.  The results are $E_m = 6005$ J/g (quail), $6413$ J/g (turkey) and
\begin{equation}\label{m3}
    \begin{split}
        \text{quail:} \qquad B(m)m &= \begin{cases} 9.53\times 10^{-4} \, m^{1.48} & m\le 75 \\
        0.00506 \, m^{1.10} & m > 75 \end{cases}\\
        \text{turkey:} \qquad B(m)m &= \begin{cases} 3.29\times 10^{-6} \, m^{1.98} & m\le 247 \\
        1.37 \times 10^{-4} \, m^{1.31} & 247 < m \le 10^{4}\\
        0.0241 \, m^{0.75} & m>10^{4}
        \end{cases}
    \end{split}
\end{equation}
where $B(m)m$ has units of W and $m$ has units of g.  The fitted theory results (solid curves in Fig.~\ref{fs2} B,D) agree very well with the $m(t)$ experimental data.  The fitted values of $E_m$ are also consistent with the range ($1400-7500$ J/g) seen for various juvenile bird species in an earlier analysis~\cite{Hou2008}.  As expected, the fraction of power input available for growth, $1 - B(m) m / \Pi(m)$, decreases as the organism develops, eventually reaching zero at the asymptotic adult mass where $B(m) m$ intersects $\Pi(m)$.

With the details of the growth model established, evaluating our expressions for $\sigma_E$ and $\sigma_B$ [Eq. (3) of the main text] and comparing them to $\sigma_E^\prime$ and $\sigma_B^\prime$ [Eq. (4) of the main text] is relatively straightforward.  Since $E(m(t)) = E_m$ is time-independent, $\sigma_E = \sigma_E^\prime =1$.  This leaves only $\sigma_B$ and $\sigma_B^\prime$, given by:
\begin{equation}\label{m4}
    \sigma_B = \langle B \rangle \langle \Theta^{-1} \rangle, \qquad \sigma^\prime_B \equiv \langle B \rangle \langle \Pi \rangle^{-1} \langle \Pi \Theta^{-1} \rangle,
\end{equation}
where the averages are given by the following integrals:
\begin{equation}\label{ints}
\begin{split}
 \langle B \rangle &= \frac{1}{t_r}\int_{m_0}^{\epsilon m_0} \frac{B(m) E_m}{\Pi(m) - B(m)m},\\
 \langle \Theta^{-1} \rangle &= \frac{1}{t_r}\int_{m_0}^{\epsilon m_0} \frac{m E_m}{(\Pi(m) - B(m)m)^2},\\
 \langle \Pi \rangle &= \frac{1}{t_r}\int_{m_0}^{\epsilon m_0} \frac{\Pi(m) E_m}{\Pi(m) - B(m)m},\\
 \langle \Pi \Theta^{-1} \rangle &= \frac{1}{t_r}\int_{m_0}^{\epsilon m_0} \frac{m \Pi(m) E_m}{(\Pi(m) - B(m)m)^2}.
\end{split}
\end{equation}
The time for reproductive maturity is $t_r = 52$ days (quail), 365 days (turkey), based on mean values from the AnAge online database~\cite{HAGR2005}.  From the theoretically fitted $m(t)$ curves this translates to $m(t_r) = \epsilon m_0$ where $\epsilon = 25$ (quail), 416 (turkey).  The integrals in Eq.~\eqref{ints} can be evaluated numerically, since all the expressions in the integrand are known based on the growth model fitting.  The results for the prefactors in Eq.~\eqref{m4} are:
\begin{equation}
    \begin{split}
        \text{quail:}\qquad \sigma_B &= 7.43, \qquad \sigma_B^\prime = 9.60\\
        \text{turkey:}\qquad \sigma_B &= 11.04, \qquad \sigma_B^\prime = 14.05
    \end{split}
\end{equation}
The discrepancy $|1 - \sigma_B^\prime/\sigma_B| = 0.27-0.29$ in both cases.  This is comparable to the simple allometric growth model (single power-law) cases investigated in the main text, where the discrepancy was always less than 50\%.  Thus the relation $s_c \approx
-\ln(R_b) \delta C_T/C_T$ continues to hold even for more complex growth models in organisms where metabolic scaling varies with developmental stage.

\section{VI. Generalizing the $s_c$ and $\tilde{s}_c$ relation for density-dependent population models}\label{dens}

The derivation in Sec.~I above, relating the baseline selection coefficient $s_c$ to the fractional change in growth rate, $\tilde{s}_c = -\delta t_r/t_r \approx \delta r/r$, can be generalized to cases where the population growth of the wild-type and mutant organisms is density-dependent.  This can occur for example when there is competition for a limited resource shared between the wild-type and mutant (i.e. a nutrient in the case of bacterial growth, or a prey population for a predatory organism), or when there are other external constraints on growth as the overall population increases.  As a result, the reproductive time $t_r$ may change from generation to generation, for example lengthening as the resource is depleted and organismal growth is slowed.  We will denote $t_r^{(k)}$ to be the mean reproductive time (duration) of the $k$th generation, and the cumulative time span of $n$ generations as $\tau_n = \sum_{k=1}^n t_r^{(k)}$, with $\tau_0 \equiv 0$.  Eq.~\eqref{s2}, defining the per-generation selection coefficient, can be adapted to this scenario as:
\begin{equation}\label{d1}
\frac{N_m(\tau_n)}{N_w(\tau_n)} = \frac{N_m(0)}{N_w(0)} \prod_{k=1}^n (1+s_c^{(k)})
\end{equation}
where we have introduced a baseline selection coefficient $s_c^{(k)}$ for the $k$th generation that can in general vary from each generation to the next.    Eq.~\eqref{d1} implies
\begin{equation}\label{d2}
\begin{split}
s_c^{(k)} = \frac{N_m(\tau_k)N_w(\tau_{k-1})}{N_w(\tau_k)N_m(\tau_{k-1})} -1 &\approx \ln \left(\frac{N_m(\tau_k)N_w(\tau_{k-1})}{N_w(\tau_k)N_m(\tau_{k-1})} \right)\\
&= \int_{\tau_{k-1}}^{\tau_k} dt\, \left(\frac{d}{dt}\ln N_m(t) - \frac{d}{dt}\ln N_w(t) \right)
\end{split}
\end{equation}
where the approximation assumes selection coefficients $|s_c^{(k)}| \ll 1$, the typical case we consider.  The integral expression for $s_c^{(k)}$ will be useful for evaluating the baseline selection coefficient later.  To make further progress, we will define a general density-dependent growth model, and subsequently illustrate it with several examples.

Let us consider wild type and mutant populations of organisms that depend on a shared resource whose quantity $S(t)$ varies in time.  Two examples where this occurs, discussed below, are 
 bacterial populations competing in a chemostat~\cite{Novick1950a,Novick1950b,Gresham2014}, and predators competing for the same prey species~\cite{Rosenzweig1963,Smith,Turchin2003}.  The resting metabolic power input has a general functional form $\Pi(m(t);S(t),N_\text{tot}(t))$, which depends on the current mass $m(t)$ of the organism, the amount of available resource $S(t)$, and potentially also on the total population $N_\text{tot}(t) = N_w(t) + N_m(t)$, i.e if there is a cooperative feeding interaction or other ecological mechanism leading to an Allee effect~\cite{Courchamp2008}.  Our focus will be on time scales covering many generations, and we will assume changes in $\Pi(m(t);S(t),N_\text{tot}(t))$ over a single generation are small enough that we can approximate $\Pi(m(t);S(t),N_\text{tot}(t)) \approx \Pi(m(t);S(\tau_{k-1}),N_\text{tot}(\tau_{k-1}))$ for $\tau_{k-1} \le t \le \tau_{k}$.  Then from Eq.~(2) of the main text we can write the reproductive time of the $k$th generation as
\begin{equation}\label{d3}
    t_r^{(k)} \approx \int_{m_0}^{\epsilon m_0} dm\,\frac{E(m)}{\Pi(m;S(\tau_{k-1}),N_\text{tot}(\tau_{k-1}))-B(m)m}
\end{equation}
for a wild-type organism with synthesis costs $E(m)$ and per-mass maintenance costs $B(m)$.  In analogy to the discussion in Sec.~I, the corresponding birth rate for the $k$th generation is $r(S(\tau_{k-1}),N_\text{tot}(\tau_{k-1})) = \ln(R_b)/t_r^{(k)}$, which now depends on $S(\tau_{k-1})$ and $N_\text{tot}(\tau_{k-1})$.  To model the population dynamics over time scales much longer than a generation, we make the usual continuum time approximation, $r(S(\tau_{k-1}),N_\text{tot}(\tau_{k-1})) \to r(S(t),N_\text{tot}(t))$, and posit a population model for $N_w(t)$ of the form
\begin{equation}\label{d4}
     \frac{dN_w}{dt} = r(S(t),N_\text{tot}(t)) N_w(t) - d N_w(t).
\end{equation}
Here $d$ is the rate at which the wild-type population is removed from the system i.e. dilution by outflow of solution in a chemostat, or death.  If $r$ was time-independent, the dynamics of $N_w(t)$ described by Eq.~\eqref{d4} would reduce to Eq.~\eqref{gr2} in Sec.~I.

We can derive an analogous equation for the mutant population $N_m(t)$.  Under our baseline assumption about the mutant organism, it has modified synthesis costs $\tilde{E}(m) = E(m) + \delta E$ and maintenance costs $\tilde{B}(m) = B(m) + \delta B$, and hence through the analogue of Eq.~\eqref{d3} it will have mean generation times $\tilde{t}_r^{(k)} = t_r^{(k)} + \delta t_r^{(k)}$ perturbed by $\delta t_r^{(k)}$ relative to the wild type.  In the continuum time population model, this translates to a mutant birth rate $\tilde{r}(S(t),N_\text{tot}(t)) = r(S(t),N_\text{tot}(t))+\delta r(S(t),N_\text{tot}(t))$, modified by some term $\delta r(S(t),N_\text{tot}(t))$, and population dynamics governed by
\begin{equation}\label{d5}
    \frac{dN_m}{dt} = \tilde{r}(S(t),N_\text{tot}(t)) N_m(t) - d N_m(t)
\end{equation}
Here $d$ is the same as in Eq.~\eqref{d4}, since we assume the mutant is removed at the same rate through dilution or other processes as the wild type.  As in the wild type case, for time-independent $r$ the solution of Eq.~\eqref{d5} for $N_m(t)$ yields the corresponding Sec.~I result, Eq.~\eqref{gr3}.

The final equation necessary to complete the description of the system is for the resource $S(t)$.  To derive it, first consider the relationship between resource consumption and metabolic expenditure.  Over the course of generation $k$, the mean resting metabolic expenditure of a wild type cell per unit time (the average input power) is given by
\begin{equation}\begin{split}\label{eqp}
    P_T^{(k)} &\approx \frac{1}{t_r^{(k)}}\int_{\tau_{k-1}}^{\tau_k} dt\, \Pi(m(t);S(\tau_{k-1}),N_\text{tot}(\tau_{k-1}))\\
    &= \frac{1}{t_r^{(k)}}\int_{m_0}^{\epsilon m_0} dm\,\frac{E(m)\Pi(m;S(\tau_{k-1}),N_\text{tot}(\tau_{k-1}))}{\Pi(m;S(\tau_{k-1}),N_\text{tot}(\tau_{k-1}))-B(m)m}.\\
    \end{split}
\end{equation}
Since this is the generalization of $\langle \Pi \rangle$ from the main text, the corresponding total resting metabolic expenditure $C_T^{(k)}$ in the $k$th generation is $C_T^{(k)} = \zeta t_r^{(k)} P_T^{(k)}$.  The mean amount of resource consumed per unit time scales with the mean input power as $\gamma^{-1} P_T^{(k)}$, where the yield parameter $\gamma$ sets the conversion rate (flux of resource needed to sustain a certain power input).  Since $\gamma^{-1} P_T^{(k)}$ depends on the substrate quantity and total population, let us denote it as a function $\gamma^{-1} P_T^{(k)} \equiv \Gamma(S(\tau_{k-1}),N_\text{tot}(\tau_{k-1}))$, the rate of resource consumption per wild type cell.  If the analogous rate for the mutant is $\tilde{\Gamma}$, then the equation for $S(t)$ takes the following form in the continuum approximation,
\begin{equation}\label{d6}
\frac{dS}{dt} = F(S(t)) -\Gamma(S(t),N_\text{tot}(t)) N_w(t) -\tilde{\Gamma}(S(t),N_\text{tot}(t))N_m(t).
\end{equation}
The function $F(S)$ describes the net production rate of the resource.

From Eqs.~\eqref{d2}, \eqref{d4}-\eqref{d5} we find:
\begin{equation}\label{sckgen}
\begin{split}
s_c^{(k)} = \int_{\tau_{k-1}}^{\tau_k} dt\, \delta r(S(t),N_\text{tot}(t))&\approx t_r^{(k)} \delta r(S(\tau_{k-1}),N_\text{tot}(\tau_{k-1}))\\
&= \ln( R_b) \frac{\delta r(S(\tau_{k-1}),N_\text{tot}(\tau_{k-1}))}{r(S(\tau_{k-1}),N_\text{tot}(\tau_{k-1}))}\\
&\equiv \ln(R_b) \tilde{s}_c^{(k)} 
\end{split}
\end{equation}
where we have used the fact that $\tau_k - \tau_{k-1} = t_r^{(k)} = \ln(R_b)/r(S(\tau_{k-1}),N_\text{tot}(\tau_{k-1}))$, and approximated the integral by assuming $\delta r(S(t),N_\text{tot}(t))$ varies slowly within a generation. We thus have a result directly analogous to the $s_c \approx \ln(R_b) \tilde{s}_c$ relation derived in Sec.~I, but now due to density dependence the fractional change in growth rate $\tilde{s}_c^{(k)}$ can in general be different for each generation $k$.

It is interesting to note that there are scenarios where $\tilde{s}_c^{(k)}$ becomes independent of $k$.  One case is when the synthesis cost is assumed independent of the organism's mass, $E(m) = E_m$, and the maintenance term is negligible $B(m) \approx 0$.  This would be a rough approximation for fast-dividing organisms like bacteria.  Using the fact that $\tilde{s}_c^{(k)}$ can also be written as $\tilde{s}_c^{(k)} = -\delta t_r^{(k)}/t_r^{(k)}$, we can see from Eq.~\eqref{d3} that a mutant with synthesis cost $E_m + \delta E$ would correspond to $\tilde{s}_c^{(k)} = -\delta E /E_m$.  In the notation of the main text, this would mean $\sigma_E=1$, and $\sigma_B$ is not defined in this case because maintenance is ignored.  Note that the $\sigma_E$ result here is independent of generation number $k$, or any specific details of the density dependence.

Another scenario where $\tilde{s}_c^{(k)}$ becomes independent of $k$ is in the long time limit for cases where the dynamical system of Eqs.~\eqref{d4}-\eqref{d5},\eqref{d6} converges to a stationary solution with non-zero total population.  In other words $N_\text{tot}(t)$ and $S(t)$ go to nonzero asymptotic values as $t \to \infty$, which we denote at $N_\text{tot}^s$ and $S^s$.  This requires that $r_\text{max}$, the maximum possible value of $r(S,N_\text{tot})$, satisfies $r_\text{max} > d$, and in addition $F(S^s) >0$.  Combinining  Eqs.~\eqref{d4}-\eqref{d5} we know that
\begin{equation}
    \frac{d}{dt}\ln \frac{N_m(t)}{N_w(t)} = \frac{1}{N_m(t)} \frac{dN_m(t)}{dt} - \frac{1}{N_w(t)} \frac{dN_w(t)}{dt} = \delta r(S(t),N_\text{tot}(t)).
\end{equation}
After an initial transient, the right-hand side relaxes to a constant value $\delta r(S^s,N_\text{tot}^s)$.  For small $\delta r$, the regime of interest in our work, we see that the ratio $N_m(t)/N_w(t)$ changes very slowly at long times. The system has two possible outcomes:  for $\delta r >0$ the ratio $N_m(t)/N_w(t)$ goes to infinity, corresponding to the mutant eventually taking over the whole population, $N_m(t) \to N_\text{tot}^s$.  The stationary state values $N^s_w$, $N^s_m$, $S^s$, satisfy relations that make the left-hand sides of Eqs.~\eqref{d4}-\eqref{d5},\eqref{d6} zero:
\begin{equation}\label{ss1}
    \tilde{r}(S^s,N_m^s) = d, \qquad N_m^s = \frac{F(S^s)}{\tilde{\Gamma}(S^s,N_m^s)}, \qquad N_w^s = 0.
\end{equation}
Alternatively when $\delta r < 0$, $N_m(t)/N_w(t)$ goes to zero, with the wild type taking over, $N_w(t) \to N_\text{tot}^s$.  Here the stationary state values satisfy:
\begin{equation}\label{ss2}
    r(S^s,N_w^s) = d, \qquad N_w^s = \frac{F(S^s)}{\Gamma(S^s,N_w^s)}, \qquad N_m^s = 0.
\end{equation}

From Eq.~\eqref{sckgen} the asymptotic relaxation dynamics at long times between mutant and wild type populations have a selection coefficient proportional to
\begin{equation}\label{scks}
\tilde{s}_c^{(k)} \approx \frac{\delta r(S^s,N_\text{tot}^s)}{r(S^s,N_\text{tot}^s)}
\end{equation}
for large $k$.  Note that Eq.~\eqref{scks} also applies if we start in the stationary state with a fully wild type population, and subsequently a mutant with small $\delta r$ appears in the population.

To illustrate the utility of Eq.~\eqref{scks}, it is instructive to consider two model systems:

\vspace{1em}
\noindent 1) {\it Bacteria in a chemostat:}  The canonical model of a bacterial populations growing in a chemostat~\cite{Gresham2014} assumes resting metabolic power input $\Pi(m;S;N_\text{tot})$ takes the form $\Pi(m;S;N_\text{tot}) = \Pi_\text{max}(m) S/(K_s + S)$, which depends on a resource $S$ but not explicitly on $N_\text{tot}$.  Here $\Pi_\text{max}(m)$ is the power input under unlimited resource, and the second term is a phenomenological hyperbolic function, first posited by Monod~\cite{Monod1949}, that expresses approximately linear dependence of the power on resource for small $S$, and saturation for $S$ much greater than some threshold value $K_s$.  We assume a simple allometric growth model with exponent $\alpha = 1$ typical of bacteria: $\Pi_\text{max}(m) = \Pi_0 m$ for some constant $\Pi_0$, $E(m) = E_m$, $B(m)=B_m$.  The integrals in Eq.~\eqref{d3} for $t_r^{(k)}$ and Eq.~\eqref{eqp} for $P_T^{(k)}$ can be evaluated directly, and the resulting birth rate $r(S(t))$ and resource consumption rate $\Gamma(S(t))$ take the form
\begin{equation}\label{r}
    r(S(t)) = \frac{1}{E_m}\left(\Pi_0 \frac{S(t)}{K_s+S(t)}-B_m\right), \qquad \Gamma(S(t)) = \frac{\Pi_0 m_0 S(t)}{\gamma (K_s + S(t))\ln 2},
\end{equation}
where we have taken $R_b = \epsilon =2$ for simplicity.  (Other values of $\epsilon$ would just add an overall prefactor.)  If $B_m$ is neglected, a common simplification for bacteria, 
$r(S)$ takes the form known as Monod's law, $r(S) = r_\text{max} S/(K_s+S)$ with a maximum rate $r_\text{max} = \Pi_0/E_m$.  However for completeness we will not ignore $B_m$, and use Eq.~\eqref{r} instead.  For typical chemostat experiments, cell death is negligible relative to $D$, the dilution rate.  This is the rate at which the solution (including the resource and bacterial cells) is removed from the system.  Hence we set $d=D$.  The final piece of the chemostat model is the net resource production, $F(S) = F_0 -DS$, where $F_0$ represents the incoming resource from a reservoir, and $-DS$ is the output flow of resource due to dilution.  The dilution rate in chemostat experiments is typically of the order of $D \sim {\cal O}(0.1)$ hr$^{-1}$~\cite{Wides2018}.

Given these model details, Eqs.~\eqref{d4}-\eqref{d5},\eqref{d6} have a a stable stationary solution with $N_\text{tot}^s >0$ in the long time limit if $D<(\Pi_0 - B_m)/E_m$.  For concreteness let us take $\delta r < 0$, so the stationary state satisfies Eq.~\eqref{ss2}.  Using Eq.~\eqref{r} to calculate $\delta r$ for perturbations to $E_m$ and $B_m$, we plug into Eq.~\eqref{scks} in the stationary limit to get
\begin{equation}
    \tilde{s}_c^{(k)} \approx -\frac{1}{E_m}\delta E_m - \frac{1}{E_m r(S^s)}\delta B_m.
\end{equation}
Hence
\begin{equation}\label{cs}
\sigma_E = 1, \qquad \sigma_B = \frac{B_m}{E_m r(S^s)} = \frac{B_m}{E_m D}.
\end{equation}
We thus see that the relative maintenance contribution $\sigma_B$ to the baseline selection coefficient can be tuned by changing $D$, making the chemostat a versatile experimental tool to explore maintenance versus synthesis cost selection pressures in bacterial populations.  As discussed in Sec.~III above, empirical data for a broad swath of unicellular organisms~\cite{Lynch2015} yields a global estimate $B_m/E_m = 3 \times 10^{-6}$ s$^{-1}$.  For a typical experimental range $D = 0.05 - 0.5$ hr$^{-1}$, we then know that $\sigma_B$ would vary between $0.22$ and $0.022$.

As noted above, the total resting metabolic expenditure in the $k$th generation $C_T^{(k)} = \zeta t_r^{(k)} P_T^{(k)}$, which can be written as $C_T^{(k)} \approx \zeta \gamma\ln(R_b)\Gamma(S^s)/r(S^s)$ for large $k$ when the system approaches the stationary state.  Using Eq.~\eqref{r} to calculate $\delta C_T^{(k)}$ for perturbations to $E_m$ and $B_m$, we find
\begin{equation}\label{tsc}
    \frac{\delta C_T^{(k)}}{C_T^{(k)}} \approx -\frac{1}{E_m}\delta E_m - \frac{1}{E_m D}\delta B_m,
\end{equation}
and so
\begin{equation}\label{tsc2}
\sigma_E^\prime = 1, \qquad \sigma_B^\prime = \frac{B_m}{E_m D}.
\end{equation}
Comparing Eqs.~\eqref{cs} and \eqref{tsc2}, we see that $\sigma_E = \sigma_E^\prime$, $\sigma_B = \sigma_B^\prime$.  Thus the approximation $\tilde{s}_c^{(k)} \approx \delta C_T^{(k)}/C_T^{(k)}$ works well in this stationary limit.  

\vspace{1em}
\noindent 2) {\it Predator-prey dynamics:}  The second example is predator-prey system where the predator wild type organism of population $N_w$ has a resting metabolic rate described by some allometric exponent $\alpha$, so that $\Pi(m;S) = \Pi_0(S) (m/m_0)^\alpha$, with a function $\Pi_0(S)$ that depends on the population of a prey organism $S$.  For convenience we have explicitly factored out $m_0^{-\alpha}$, in order that $\Pi_0(S)$ have units of W.  As in the allometric model of the main text, $E(m) = E_m$, $B(m) = B_m$.  From Eq.~\eqref{d3} the predator reproductive time for $\alpha \ne 1$ is given by:
\begin{equation}\label{p1}
\begin{split}
    t_r^{(k)} &= \frac{E_m}{B_m(1-\alpha)} \ln \left(\frac{\Pi _0(S(t))-m_0 B_m}{\Pi _0(S(t))-m_0 \epsilon ^{1-\alpha } B_m}\right).
    \end{split}
\end{equation}
The population growth rate $r(S(\tau_{k-1})) = \ln( R_b)/t_r^{(k)}$.  Since the exponent $\alpha$ for metazoans is typically close to 1 (as in the common choice of $\alpha =3/4$~\cite{West2001,Hou2008}), we will simplify the expression for $r(S(\tau_{k-1}))$ by expanding it to first order in $\alpha$ around $\alpha =1$.  In the continuum version, where $S(\tau_{k-1}) \to S(t)$, the growth rate then takes the form
\begin{equation}\label{p2}
r(S(t)) \approx \mu \Pi_0(S(t)) - \nu,
\end{equation}
where
\begin{equation}\label{p3}
\mu \equiv \frac{ \left[2- (1-\alpha) \epsilon\right] \ln R_b}{2 E_m m_0 \ln \epsilon}, \qquad \nu \equiv \frac{B_m \ln R_b}{E_m \ln \epsilon}.
\end{equation}
This approximation works best when $|1-\alpha|\ln \epsilon \ll 2$.  The first term in the $r(S(t))$ expression is the contribution to population growth of metabolic power input through prey consumption, while the second term describes the degree to which the growth rate is reduced by having some of that power channeled into maintenance.  We approximate Eq.~\eqref{eqp} for $P_T^{(k)}$ to lowest order in $1-\alpha$, giving the following expression for the resource consumption rate,
\begin{equation}\label{p3b}
    \Gamma(S(t)) \approx \frac{(\epsilon -1)\Pi_0(S(t))}{\gamma \ln \epsilon}.
\end{equation}

Eqs.~\eqref{p2}-\eqref{p3b} can be plugged into the dynamical system of Eqs.~\eqref{d4}-\eqref{d5},\eqref{d6}.  To complete the description of the predator-prey system, we make several additional assumptions:  we interpret $d$ as the death rate of the predator, and choose a prey reproduction function $F(S) = b S (1-S/K)$, where $b$ is the maximum net rate at which the prey species reproduces itself, $K$ is the prey carrying capacity.  Finally we choose $\Pi_0(S) = \Pi_0^\text{max} S/(S_c+S)$ for some constants $\Pi_0^\text{max}$ and $S_c$, where $\Pi_0^\text{max}$ is the maximum metabolic power input in the limit of unlimited prey abundance, and $S_c$ is the critical prey population above which saturation of $\Pi_0(S)$ occurs.  This has the form of a Holling type II predator-prey functional response~\cite{Holling1959,Turchin2003}, which assumes that when prey becomes overabundant, the consumption rate is limited by the finite time required for the predator to process each kill, hence leading to saturation of $\Pi_0(S)$.  Given the above assumptions, the dynamical system of Eqs.~\eqref{d4}, \eqref{d6} reduces to the Rosenzweig-Macarthur (RM) predator-prey model~\cite{Rosenzweig1963,Smith} in the absence of a mutant predator population, and with maintenance $B_m$ neglected.  Here we consider the generalized RM model with $B_m>0$ and with populations of both wild type and mutant predators competing for the same prey.

To apply Eq.~\eqref{scks}, we need to establish the properties of the stationary solution of Eqs.~\eqref{d4}-\eqref{d5},\eqref{d6}, along the lines of the stability analysis for the original RM model in Ref.~\cite{Smith}.  For concreteness let us assume $\delta r < 0$, so the stationary solution should satisfy Eq.~\eqref{ss2}.  Then a stationary solution exists with $N_\text{tot}^s >0$ if $d < \mu\Pi_0^\text{max}K(K+S_c)^{-1} - \nu$, and takes the form
\begin{equation}
    N^s_w = \frac{b \gamma \mu S^s (K-S^s)\ln \epsilon}{(\epsilon-1) K (d+\nu)}, \qquad N^s_m = 0, \qquad S^s = \frac{S_c(d+\nu)}{\mu \Pi_0^\text{max} -\nu -d}.
\end{equation}
This solution is stable when $S^s > (K-S_c)/2$, which will be the parameter regime on which we focus.  When the stationary solution is unstable, the system exhibits limit cycles, which are beyond the scope of this analysis (though in such cases Eq.~\eqref{sckgen} would still hold, with an $s_c^{(k)}$ that becomes periodic in $k$ in the long time limit).

Putting everything together, we can use Eqs.~\eqref{p2}-\eqref{p3} to calculate $\delta r$ for perturbations to $E_m$ and $B_m$, and we then evaluate Eq.~\eqref{scks} at the stationary state,
\begin{equation}
    \tilde{s}_c^{(k)} \approx -\frac{1}{E_m}\delta E_m - \frac{1}{E_m d}\left(\frac{\ln  R_b}{\ln \epsilon}\right)\delta B_m.
\end{equation}
Thus in this case,
\begin{equation}
\sigma_E = 1, \qquad \sigma_B = \frac{B_m}{E_m d}\left( \frac{\ln  R_b}{ \ln \epsilon}\right).
\end{equation}
The final result for $\sigma_B$ has a very similar form to the bacterial chemostat example above, Eq.~\eqref{cs}, except for an additional factor of $\ln R_b / \ln \epsilon$, and $d$ being the predator death rate, rather than a dilution rate.  As discussed in Sec.~III above, the ratio $B_m/E_m$ is expected to be somewhat different for multicellular versus unicellular species, with for example $B_m/E_m = 10^{-7} - 10^{-6}$ s$^{-1}$ for a range of mammals~\cite{Hou2008}.  The scale of the dimensionless factor  $\ln  R_b / \ln \epsilon$ can also be estimated. For example, data from a variety of fissiped carnivore and insectivore mammals yields a range $\ln  R_b / \ln \epsilon \approx 0.2 - 0.4$~\cite{Sibly2009}.  Given these estimates, we will get $\sigma_B > \sigma_E =1$ when the mean predator lifetime $d^{-1}$ is on the order of a year or higher.  If we do the analogous $C_T^{(k)}$ perturbation analysis as in the previous example, we again find that $\sigma_E = \sigma_E^\prime$, $\sigma_B = \sigma_B^\prime$, validating the approximation $\tilde{s}_c^{(k)} \approx \delta C_T^{(k)}/C_T^{(k)}$ for large $k$.

\section{VII. Understanding the roles of baseline metabolic costs versus adaptive effects in selection}\label{adap}

In the main text our focus has been on understanding the baseline contribution to selection, $s_c$, arising from changes in metabolic costs (perturbations to synthesis and maintenance).  However in the general case a genetic variant will have additional perturbations to its phenotype contributing to the selection coefficient, yielding an overall coefficient $s = s_c + s_a$ with an adaptive correction $s_a$~\cite{Lynch2015}.  In this section we illustrate how our bioenergetic growth formalism can be extended to consider perturbations beyond metabolic costs, yielding expressions for $s_a$.  We show that even with these additional perturbations, $s_c$ retains the form of our original formalism, and can be generally related to associated fractional changes in the total resting metabolic expenditure $C_T$.  However the relationship between $s_a$ and $C_T$ is more complicated, dependent on system-specific details of the organism and its environment.  Thus investigating the relative magnitude of $s_c$ and $s_a$ in individual biological cases, and the ways in which $s_a$ might or might not compensate for the metabolic costs encapsulated in $s_c$, opens a rich set of questions for future study.

We proceed by generalizing the derivations in SI Secs.~\ref{sc} and \ref{der} to include two examples of adaptive effects for the mutant organisms:  changes in the death rate $d$, and changes in resting metabolic power input $\Pi(m(t))$.  These are not the only quantities in the theory that could potentially be subject to adaptation:  other possibilities include the the mean number of offspring $R_b$ or the reproductive maturity parameter $\epsilon$.  However these two examples illustrate the general scheme for how to include adaptive effects in the formalism.  A more comprehensive treatment of the interplay between baseline metabolic costs and adaptive effects in selection is the subject of an upcoming work~\cite{Ilker2019}.

In the original derivation of SI Sec.~I, where we considered only the baseline selective contribution due to metabolic costs, we assumed the mutant has the same mean number of offspring $R_b$ and death rate $d= -\ln(p(t_r))/t_r$ as the wild type.  We now partially relax that assumption:  we still keep $R_b$ the same, but allow the mutant death rate $\tilde{d} = -\ln(p(t_r+\delta t_r))/(t_r+\delta t_r)$ to be different than $d$ by some amount $\delta d \equiv \tilde{d} - d$.  For example if the survival probability $p(t)$ was exponential for the wild-type, $p(t) = \exp(-\eta t)$, and the rate $\eta$ was modified in the mutant to $\tilde{\eta}$, then $\delta d = \tilde{\eta} - \eta$.  One case where such considerations would come into play would be bacterial strains growing in an environment with antibiotics:  a variant that acquired antibiotic resistance would have a lowered death rate, $\delta d < 0$, but also possibly non-trivial metabolic costs associated with maintaining the resistance~\cite{Dahlberg2003,Melnyk2015,Zampieri2017}, as discussed in more detail below.  Since the mutant now has both altered birth rate $\tilde{r} = r+\delta r = \ln(R_b)/(t_r+\delta t_r)$ and death rate $\tilde{d}$, the population growth equation [Eq.~\eqref{gr3}] becomes
\begin{equation}\label{a1}
N_m(t) = e^{(\tilde r - \tilde d)t}N_m(0),
\end{equation}
and the ratio of mutant to wild-type populations [Eq.~\eqref{s1}] is changed to
\begin{equation}\label{a2}
\frac{N_m (t)}{N_w (t)}= \frac{N_m (0)}{N_w (0)}e^{t (\delta r-\delta d)} = \frac{N_m (0)}{N_w (0)}R_b ^{t\left (\frac{1}{t_{r}+\delta t_r}-\frac{1}{t_{r}} \right)}e^{-t \delta d}.
\end{equation}
Since we now have both metabolic costs and adaptive effects, Eq.~\eqref{s2} now defines the full selection coefficient $s$ instead of just $s_c$, evaluated after $n$ wild-type generations, $t = nt_r$,
\begin{equation}\label{a3}
\frac{N_m (n t_r)}{N_w (n t_r)}=\frac{N_m (0)}{N_w (0)}(1+s)^{n}.
\end{equation}
Comparing Eq.~\eqref{a3} to Eq.~\eqref{a2}, we can write
\begin{equation}\label{a4}
    s = R_b^{-\frac{\delta t_r}{t_r+\delta t_r}}e^{-t_r \delta d} -1 = R_b^{\frac{\tilde{s}}{1-\tilde{s}}}e^{-t_r \delta d} -1,
\end{equation}
where $\tilde{s} \equiv -\delta t_r/t_r \approx \delta r /r$, with the latter approximation valid when $|\tilde{s}| \ll 1$.  Note that in this general case $\delta r$ (or equivalently $\delta t_r$) describes the perturbation to the growth rate both from the altered metabolic costs as well as adaptive effects (as we will see below when we consider changes in $\Pi(m(t))$).  We will distinguish the metabolic versus adaptive contributions shortly.  If the magnitudes of $\delta d$ and $\delta t_r$ are small relative to $t_r$, Eq.~\eqref{a4} to first order in the perturbations $\delta d$ and $\delta t_r$ is given by
\begin{equation}\label{a5}
    s \approx -t_r \delta d + \ln(R_b) \tilde{s}.
\end{equation}
To complete the description of $s$, we now would like to expand out the $\tilde{s}$ contribution on the right-hand side of Eq.~\eqref{a5} into contributions from metabolic and adaptive perturbations, generalizing the derivation of SI Sec.~II for $\tilde{s}_c$.  Starting with Eq.~\eqref{dd1} for $t_r$, we consider how $t_r$ will be altered under the simultaneous perturbations: $E(m) \to E(m) +\delta E$, $B(m) \to B(m) + \delta B$, $\Pi(m) \to \Pi(m) +\delta \Pi(m)$.  The first two are the baseline metabolic changes in synthesis and maintenance costs we considered before.  The last one represents the possibility that the genetic variant could also have modifications in the resting metabolic power input function $\Pi(m(t))$.  As before we assume $\delta E$ and $\delta B$ are independent of $m(t)$ for simplicity, but we allow the power input perturbation $\delta \Pi(m(t))$ to have some general $m(t)$-dependent functional form.  Eq.~\eqref{dd2} then becomes
\begin{equation}\label{a6}
\begin{split}
t_r + \delta t_r &= \int_{m_0}^{\epsilon m_0} dm \frac{E(m)+\delta E}{(\Pi(m)+\delta \Pi(m)) - (B(m)+\delta B) m}\\
&\approx t_r + \delta E \int_{m_0}^{\epsilon m_0} dm \frac{1}{G(m)} + \delta B \int_{m_0}^{\epsilon m_0} dm \frac{ E(m)}{\Theta(m) G(m)} - \int_{m_0}^{\epsilon m_0} dm\,\frac{E(m)\delta \Pi(m) }{G^2(m)},
\end{split}
\end{equation}
where we have expanded to first order in the perturbations, and $G(m) = \Pi(m) - B(m)m$, $\Theta(m) = G(m)/m$.  As in SI Sec.~II, we can rewrite the perturbation terms on the right-hand side of Eq.~\eqref{a6} in terms of averages over $p(m) = t_r^{-1} E(m)/G(m)$, and get an expression for $\tilde{s}$,
\begin{equation}\label{a7}
\begin{split}
    \tilde{s} &= -\langle E^{-1} \rangle \delta E - \langle \Theta^{-1} \rangle \delta B + \langle G^{-1} \delta \Pi \rangle\\
    &= \underbrace{-\sigma_E\frac{\delta E}{\langle E \rangle} - \sigma_B \frac{\delta B}{\langle B \rangle}}_{\tilde{s}_c} + \underbrace{\langle G^{-1} \delta \Pi\rangle}_{\tilde{s}_a},
\end{split}
\end{equation}
where the synthesis/maintenance prefactors have the same definition as before: $\sigma_E = \langle E \rangle \langle E^{-1}\rangle$, $\sigma_B = \langle B \rangle \langle \Theta^{-1} \rangle$.  The first two terms of Eq.~\eqref{a7} are thus $\tilde{s}_c$ [Eq.~\eqref{dd5}], the baseline contribution to $\tilde{s}$ from perturbations to metabolic synthesis/maintenance costs.  The new adaptive contribution $\tilde{s}_a = \langle G^{-1} \delta \Pi \rangle$ comes from the perturbation to the power input, and has a straightforward physical interpretation:  since $G(m)$ is the power available for growth (once maintenance costs are subtracted away from the power input), the average $\langle G^{-1}\delta \Pi\rangle$ is just the fractional size of the power input perturbation $\delta \Pi(m)$  relative to $G(m)$, averaged over the course of a generation.  Plugging Eq.~\eqref{a7} into Eq.~\eqref{a5}, we get our final expression for $s$,
\begin{equation}\label{a8}
    s \approx \underbrace{\ln(R_b)\tilde{s}_c}_{s_c} +\underbrace{\ln(R_b) \tilde{s}_a -t_r \delta d }_{s_a}.
\end{equation}
The first term is $s_c$, what we had calculated originally under the baseline assumption, and the remaining terms we can identify as $s_a$, the correction due to adaptive effects beyond the baseline assumption.

As we did for the baseline case, we can compare the contributions to $\tilde{s}$ in Eq.~\eqref{a7} from the different perturbations to the associated changes in the total resting metabolic expenditure per generation, $C_T = \zeta t_r \langle \Pi \rangle$.  Again proceeding analogously to SI Sec.~II, Eqs.~\eqref{dd7}-\eqref{dd11}, we find the fractional change 
\begin{equation}\label{a9}
\frac{\delta C_T}{C_T} = \underbrace{\sigma_E^\prime \frac{\delta E}{\langle E \rangle} +\sigma_B^\prime \frac{\delta B}{\langle B \rangle}}_{\equiv \left(\frac{\delta C_T}{C_T}\right)_c} + \underbrace{\langle \Pi \rangle^{-1} \langle \delta \Pi \rangle - \langle \Pi \rangle^{-1} \langle \Pi G^{-1}\delta \Pi\rangle}_{\equiv \left(\frac{\delta C_T}{C_T}\right)_a},
\end{equation}
where we have split $\delta C_T/C_T$ into the baseline cost contribution $(\delta C_T/C_T)_c$ and the adaptive correction $(\delta C_T/C_T)_a$.  Here $\sigma^\prime_E$, $\sigma^\prime_B$ have the same definitions as before, given in Eq.~\eqref{dd11}.  As we argued in the main text, the prefactors $\sigma_E^\prime$, $\sigma_B^\prime$ are comparable to $\sigma_E$, $\sigma_B$ respectively across a wide range of biologically relevant growth scenarios. Thus the maintenance/synthesis contributions to $\tilde{s}$ in Eq.~\eqref{a7} have a direct reflection in the baseline contribution to $\delta C_T/C_T$ in Eq.~\eqref{a9}.  In other words $\tilde{s}_c \approx - (\delta C_T/C_T)_c$, the baseline relation that is the central focus of the main text.

\begin{figure}
    \centering
    \includegraphics[width=0.67\textwidth]{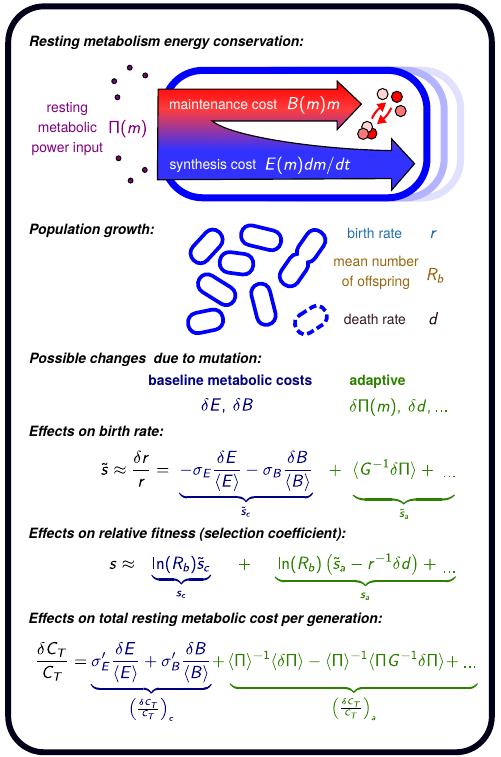}
    \caption{\linespread{1.0}\selectfont{} Schematic summary of the basic theoretical formalism. Energy conservation for individual organisms implies a growth relation for mass $m(t)$:  resting metabolic power input $\Pi(m)$ is expended in maintenance $B(m) m$ and synthesis $E(m) dm/dt$ [main text Eq.~(1)].  The overall population has a birth rate $r = \ln(R_b)/t_r$ and death rate $d$, where $R_b$ is the mean number of offspring and $t_r$ is the mean generation time [related to metabolism through main text Eq.~(2)].  Mutations can introduce perturbations in the baseline metabolic costs $\delta E$, $\delta B$ as well as adaptive effects like $\delta \Pi(m)$, $\delta d$.  These have several consequences: i) modifying the generation time by $\delta t_r$ [SI Eq.~60], which in turn changes the birth rate by $\delta r$ and leads to a relative fitness difference quantified by the selection coefficient $s$; ii) modifying the total resting metabolic expenditure $C_T$ over a generation by an amount $\delta C_T$.  All these changes can be expressed in terms of time averages of metabolic quantities integrated over the wild type generation time $t_r$, i.e. $\langle F \rangle \equiv t_r^{-1} \int_0^{t_r} dt F (m(t))$.  The definitions of $\sigma_E, \sigma_B$, $\sigma^\prime_E, \sigma^\prime_B$ are given in main text Eqs.~(3-4), and $G(m) \equiv \Pi(m) - B(m)m \ge 0$ is the power channeled to growth.  The baseline metabolic contributions due to $\delta E$ and $\delta B$ we denote with subscript $c$, and the remaining (adaptive) contributions with subscript $a$.}
    \label{scheme}
\end{figure}

For the adaptive components, the relation between $\tilde{s}_a$ from Eq.~\eqref{a7} and $(\delta C_T/C_T)_a$ from Eq.~\eqref{a9} is not so straightforward.  Moreover the full adaptive contribution to the selection coefficient, $s_a$ from Eq.~\eqref{a8}, includes a term reflecting the change in death rate, $-t_r \delta d$, which has no analogue in $(\delta C_T/C_T)_a$.  As a concrete example, consider the allometric growth model described in the main text, where $\Pi(m(t)) = \Pi_0 m^\alpha(t)$, $E(m(t)) = E_m$, $B(m(t)) = B_m$.  Let us focus on the simplest case, $\alpha =1$, corresponding to exponential cell mass growth, where the baseline relation is exact, $\tilde{s}_c = - (\delta C_T/C_T)_c$ since $\sigma_E = \sigma_E^\prime =1$ and $\sigma_B = \sigma_B^\prime$.  If the power input perturbation occurs through the prefactor, $\Pi_0 \to \Pi_0 + \delta \Pi_0$, then $\delta \Pi(m) = \delta \Pi_0 m$.  We can then evaluate both $\tilde{s}_a$ and $(\delta C_T/C_T)_a$:
\begin{equation}\label{a10}
\begin{split}
\tilde{s}_a &= \langle G^{-1}\delta \Pi \rangle = \frac{\delta \Pi_0}{\Pi_0 - B_m},\\
\left(\frac{\delta C_T}{C_T}\right)_a &= \langle \Pi \rangle^{-1} \langle \delta \Pi \rangle - \langle \Pi \rangle^{-1} \langle \Pi G^{-1}\delta \Pi\rangle = \frac{\delta \Pi_0} {\Pi_0} - \frac{\delta \Pi_0}{\Pi_0 - B_m}.
\end{split}
\end{equation}
So we see that even in this simple case $\tilde{s}_a$ and $(\delta C_T/C_T)_a$ differ by an extra term in the latter.  The significance of this difference depends on the details of the specific system under consideration.  A schematic summary of the above results is shown in Fig.~\ref{scheme}.

The relative contributions of $s_c$ and $s_a$ to selection in particular biological systems may also depend on aspects of the environment, with $s_c$ or $s_a$ playing a leading role in different contexts.  For bacterial strains evolving in the presence of an antibiotic, the adaptive benefit of a mutation that confers antibiotic resistance (i.e. lowering the death rate $d$ in the example above) will be reflected in a significant positive fitness contribution $s_a > 0$.  But the molecular mechanisms that lead to resistance typically come with non-negligible metabolic costs ($s_c<0$) that slow bacterial growth in the absence of drug, for example over-expression of energy-consuming multi-drug efflux pumps~\cite{Blanco2016}, or switching to less energy-efficient biochemical pathways that are not targeted by the drug~\cite{Zampieri2017}.  While $s_a$ may more than compensate for $s_c$ when the antibiotic is present, the resistant strains are at least initially at a disadvantage in environments without the antibiotic, where $s_c$ is likely the dominant contribution to $s$.  Ref.~\cite{Melnyk2015} conducted meta-analysis of competitive fitness assays between wild type and resistant bacteria in drug-free conditions, focusing on cases where the resistance was conferred by a single mutation.  They found mean values of $s$ for mutants resistant to different drugs to range between $\approx -0.03$ to $-0.28$ for eight different bacterial species.  Given the large effective populations of bacteria, these values are large enough to put the mutant strains under strong selective pressure to reduce the metabolic price of maintaining resistance.  Intriguingly the net result of this pressure under subsequent drug-free evolution is typically not the loss of the resistance, but rather additional mutations that compensate for the metabolic costs~\cite{Dahlberg2003,Melnyk2015,Zampieri2017}.

Another scenario where environmental conditions can determine the significance of metabolic costs to selection is for systems that exhibit spare respiratory capacity:  an energetic reserve that allows them to maintain ATP levels to cope with stress that increases energy demand~\cite{Choi2009,Brand2011,Nickens2013,Pfleger2015}.  While this is particularly relevant for multicellular eukaryotes, it may also be facilitated in prokaryotic populations by large cell-to-cell variations in ATP concentrations~\cite{Yaginuma2014}.  If such spare capacity exists, then under unstressed conditions a mutation with extra baseline metabolic costs may get absorbed by the reserve, with these costs not translating into a penalty in terms of cell growth rate (up to a certain cost threshold).  In other words, after factoring in the spare capacity the effective $\delta E$ and $\delta B$ of this mutation would have a smaller magnitude (or be zero) compared to the case where no spare capacity existed, and hence the magnitude of $s_c$ would also be smaller. However these costs would not necessarily always remain hidden from selection:  if the environment switched to stress conditions with higher energy demands over the long term, these mutants would have a lower reserve capacity to cope.  In this case there would be less buffering of the costs, and a larger magnitude of $s_c$ for the mutants.

In summary, quantitative understanding of how $s$ splits into $s_c$ and $s_a$ components, and their relative importance under different conditions, will be very useful in the probing the metabolic influences on evolution in biological systems, and an important target for future work~\cite{Ilker2019}.


\begin{thebibliography}{81}%
\makeatletter
\providecommand \@ifxundefined [1]{%
 \@ifx{#1\undefined}
}%
\providecommand \@ifnum [1]{%
 \ifnum #1\expandafter \@firstoftwo
 \else \expandafter \@secondoftwo
 \fi
}%
\providecommand \@ifx [1]{%
 \ifx #1\expandafter \@firstoftwo
 \else \expandafter \@secondoftwo
 \fi
}%
\providecommand \natexlab [1]{#1}%
\providecommand \enquote  [1]{``#1''}%
\providecommand \bibnamefont  [1]{#1}%
\providecommand \bibfnamefont [1]{#1}%
\providecommand \citenamefont [1]{#1}%
\providecommand \href@noop [0]{\@secondoftwo}%
\providecommand \href [0]{\begingroup \@sanitize@url \@href}%
\providecommand \@href[1]{\@@startlink{#1}\@@href}%
\providecommand \@@href[1]{\endgroup#1\@@endlink}%
\providecommand \@sanitize@url [0]{\catcode `\\12\catcode `\$12\catcode
  `\&12\catcode `\#12\catcode `\^12\catcode `\_12\catcode `\%12\relax}%
\providecommand \@@startlink[1]{}%
\providecommand \@@endlink[0]{}%
\providecommand \url  [0]{\begingroup\@sanitize@url \@url }%
\providecommand \@url [1]{\endgroup\@href {#1}{\urlprefix }}%
\providecommand \urlprefix  [0]{URL }%
\providecommand \Eprint [0]{\href }%
\providecommand \doibase [0]{http://dx.doi.org/}%
\providecommand \selectlanguage [0]{\@gobble}%
\providecommand \bibinfo  [0]{\@secondoftwo}%
\providecommand \bibfield  [0]{\@secondoftwo}%
\providecommand \translation [1]{[#1]}%
\providecommand \BibitemOpen [0]{}%
\providecommand \bibitemStop [0]{}%
\providecommand \bibitemNoStop [0]{.\EOS\space}%
\providecommand \EOS [0]{\spacefactor3000\relax}%
\providecommand \BibitemShut  [1]{\csname bibitem#1\endcsname}%
\let\auto@bib@innerbib\@empty
\bibitem [{\citenamefont {Dekel}\ and\ \citenamefont {Alon}(2005)}]{Dekel2005}%
  \BibitemOpen
  \bibfield  {author} {\bibinfo {author} {\bibfnamefont {E.}~\bibnamefont
  {Dekel}}\ and\ \bibinfo {author} {\bibfnamefont {U.}~\bibnamefont {Alon}},\
  }\href@noop {} {\bibfield  {journal} {\bibinfo  {journal} {Nature}\ }\textbf
  {\bibinfo {volume} {436}},\ \bibinfo {pages} {588} (\bibinfo {year}
  {2005})}\BibitemShut {NoStop}%
\bibitem [{\citenamefont {Dill}\ \emph {et~al.}(2011)\citenamefont {Dill},
  \citenamefont {Ghosh},\ and\ \citenamefont {Schmit}}]{dill2011physical}%
  \BibitemOpen
  \bibfield  {author} {\bibinfo {author} {\bibfnamefont {K.~A.}\ \bibnamefont
  {Dill}}, \bibinfo {author} {\bibfnamefont {K.}~\bibnamefont {Ghosh}}, \ and\
  \bibinfo {author} {\bibfnamefont {J.~D.}\ \bibnamefont {Schmit}},\
  }\href@noop {} {\bibfield  {journal} {\bibinfo  {journal} {Proc. Natl. Acad.
  Sci.}\ }\textbf {\bibinfo {volume} {108}},\ \bibinfo {pages} {17876}
  (\bibinfo {year} {2011})}\BibitemShut {NoStop}%
\bibitem [{\citenamefont {ten Wolde}\ \emph {et~al.}(2016)\citenamefont {ten
  Wolde}, \citenamefont {Becker}, \citenamefont {Ouldridge},\ and\
  \citenamefont {Mugler}}]{ten2016fundamental}%
  \BibitemOpen
  \bibfield  {author} {\bibinfo {author} {\bibfnamefont {P.~R.}\ \bibnamefont
  {ten Wolde}}, \bibinfo {author} {\bibfnamefont {N.~B.}\ \bibnamefont
  {Becker}}, \bibinfo {author} {\bibfnamefont {T.~E.}\ \bibnamefont
  {Ouldridge}}, \ and\ \bibinfo {author} {\bibfnamefont {A.}~\bibnamefont
  {Mugler}},\ }\href@noop {} {\bibfield  {journal} {\bibinfo  {journal} {J.
  Stat. Phys.}\ }\textbf {\bibinfo {volume} {162}},\ \bibinfo {pages} {1395}
  (\bibinfo {year} {2016})}\BibitemShut {NoStop}%
\bibitem [{\citenamefont {Hathcock}\ \emph {et~al.}(2016)\citenamefont
  {Hathcock}, \citenamefont {Sheehy}, \citenamefont {Weisenberger},
  \citenamefont {Ilker},\ and\ \citenamefont {Hinczewski}}]{hathcock2016noise}%
  \BibitemOpen
  \bibfield  {author} {\bibinfo {author} {\bibfnamefont {D.}~\bibnamefont
  {Hathcock}}, \bibinfo {author} {\bibfnamefont {J.}~\bibnamefont {Sheehy}},
  \bibinfo {author} {\bibfnamefont {C.}~\bibnamefont {Weisenberger}}, \bibinfo
  {author} {\bibfnamefont {E.}~\bibnamefont {Ilker}}, \ and\ \bibinfo {author}
  {\bibfnamefont {M.}~\bibnamefont {Hinczewski}},\ }\href@noop {} {\bibfield
  {journal} {\bibinfo  {journal} {IEEE Trans. Mol. Biol. Multi-Scale Commun.}\
  }\textbf {\bibinfo {volume} {2}},\ \bibinfo {pages} {16} (\bibinfo {year}
  {2016})}\BibitemShut {NoStop}%
\bibitem [{\citenamefont {Zechner}\ \emph {et~al.}(2016)\citenamefont
  {Zechner}, \citenamefont {Seelig}, \citenamefont {Rullan},\ and\
  \citenamefont {Khammash}}]{zechner2016molecular}%
  \BibitemOpen
  \bibfield  {author} {\bibinfo {author} {\bibfnamefont {C.}~\bibnamefont
  {Zechner}}, \bibinfo {author} {\bibfnamefont {G.}~\bibnamefont {Seelig}},
  \bibinfo {author} {\bibfnamefont {M.}~\bibnamefont {Rullan}}, \ and\ \bibinfo
  {author} {\bibfnamefont {M.}~\bibnamefont {Khammash}},\ }\href@noop {}
  {\bibfield  {journal} {\bibinfo  {journal} {Proc. Natl. Acad. Sci.}\ }\textbf
  {\bibinfo {volume} {113}},\ \bibinfo {pages} {4729} (\bibinfo {year}
  {2016})}\BibitemShut {NoStop}%
\bibitem [{\citenamefont {Fancher}\ and\ \citenamefont
  {Mugler}(2017)}]{fancher2017fundamental}%
  \BibitemOpen
  \bibfield  {author} {\bibinfo {author} {\bibfnamefont {S.}~\bibnamefont
  {Fancher}}\ and\ \bibinfo {author} {\bibfnamefont {A.}~\bibnamefont
  {Mugler}},\ }\href@noop {} {\bibfield  {journal} {\bibinfo  {journal} {Phys.
  Rev. Lett.}\ }\textbf {\bibinfo {volume} {118}},\ \bibinfo {pages} {078101}
  (\bibinfo {year} {2017})}\BibitemShut {NoStop}%
\bibitem [{\citenamefont {Kimura}(1962)}]{Kimura1962}%
  \BibitemOpen
  \bibfield  {author} {\bibinfo {author} {\bibfnamefont {M.}~\bibnamefont
  {Kimura}},\ }\href@noop {} {\bibfield  {journal} {\bibinfo  {journal}
  {Genetics}\ }\textbf {\bibinfo {volume} {47}},\ \bibinfo {pages} {713}
  (\bibinfo {year} {1962})}\BibitemShut {NoStop}%
\bibitem [{\citenamefont {Ohta}(1973)}]{Ohta1973}%
  \BibitemOpen
  \bibfield  {author} {\bibinfo {author} {\bibfnamefont {T.}~\bibnamefont
  {Ohta}},\ }\href@noop {} {\bibfield  {journal} {\bibinfo  {journal} {Nature}\
  }\textbf {\bibinfo {volume} {246}},\ \bibinfo {pages} {96} (\bibinfo {year}
  {1973})}\BibitemShut {NoStop}%
\bibitem [{\citenamefont {Ohta}\ and\ \citenamefont
  {Gillespie}(1996)}]{Ohta1996}%
  \BibitemOpen
  \bibfield  {author} {\bibinfo {author} {\bibfnamefont {T.}~\bibnamefont
  {Ohta}}\ and\ \bibinfo {author} {\bibfnamefont {J.~H.}\ \bibnamefont
  {Gillespie}},\ }\href@noop {} {\bibfield  {journal} {\bibinfo  {journal}
  {Theor. Popul. Biol.}\ }\textbf {\bibinfo {volume} {49}},\ \bibinfo {pages}
  {128} (\bibinfo {year} {1996})}\BibitemShut {NoStop}%
\bibitem [{\citenamefont {Orgel}\ and\ \citenamefont
  {Crick}(1980)}]{Orgel1980}%
  \BibitemOpen
  \bibfield  {author} {\bibinfo {author} {\bibfnamefont {L.~E.}\ \bibnamefont
  {Orgel}}\ and\ \bibinfo {author} {\bibfnamefont {F.~H.}\ \bibnamefont
  {Crick}},\ }\href@noop {} {\bibfield  {journal} {\bibinfo  {journal}
  {Nature}\ }\textbf {\bibinfo {volume} {284}},\ \bibinfo {pages} {604}
  (\bibinfo {year} {1980})}\BibitemShut {NoStop}%
\bibitem [{\citenamefont {Wagner}(2005)}]{Wagner2005}%
  \BibitemOpen
  \bibfield  {author} {\bibinfo {author} {\bibfnamefont {A.}~\bibnamefont
  {Wagner}},\ }\href@noop {} {\bibfield  {journal} {\bibinfo  {journal} {Mol.
  Biol. Evol.}\ }\textbf {\bibinfo {volume} {22}},\ \bibinfo {pages} {1365}
  (\bibinfo {year} {2005})}\BibitemShut {NoStop}%
\bibitem [{\citenamefont {Lynch}\ and\ \citenamefont
  {Marinov}(2015)}]{Lynch2015}%
  \BibitemOpen
  \bibfield  {author} {\bibinfo {author} {\bibfnamefont {M.}~\bibnamefont
  {Lynch}}\ and\ \bibinfo {author} {\bibfnamefont {G.~K.}\ \bibnamefont
  {Marinov}},\ }\href@noop {} {\bibfield  {journal} {\bibinfo  {journal} {Proc.
  Natl. Acad. Sci.}\ }\textbf {\bibinfo {volume} {112}},\ \bibinfo {pages}
  {15690} (\bibinfo {year} {2015})}\BibitemShut {NoStop}%
\bibitem [{\citenamefont {Gillespie}(2010)}]{Gillespie2010}%
  \BibitemOpen
  \bibfield  {author} {\bibinfo {author} {\bibfnamefont {J.~H.}\ \bibnamefont
  {Gillespie}},\ }\href@noop {} {\emph {\bibinfo {title} {Population genetics:
  a concise guide}}}\ (\bibinfo  {publisher} {JHU Press},\ \bibinfo {year}
  {2010})\BibitemShut {NoStop}%
\bibitem [{\citenamefont {Sung}\ \emph {et~al.}(2012)\citenamefont {Sung},
  \citenamefont {Ackerman}, \citenamefont {Miller}, \citenamefont {Doak},\ and\
  \citenamefont {Lynch}}]{Sung2012}%
  \BibitemOpen
  \bibfield  {author} {\bibinfo {author} {\bibfnamefont {W.}~\bibnamefont
  {Sung}}, \bibinfo {author} {\bibfnamefont {M.~S.}\ \bibnamefont {Ackerman}},
  \bibinfo {author} {\bibfnamefont {S.~F.}\ \bibnamefont {Miller}}, \bibinfo
  {author} {\bibfnamefont {T.~G.}\ \bibnamefont {Doak}}, \ and\ \bibinfo
  {author} {\bibfnamefont {M.}~\bibnamefont {Lynch}},\ }\href@noop {}
  {\bibfield  {journal} {\bibinfo  {journal} {Proc. Natl. Acad. Sci.}\ }\textbf
  {\bibinfo {volume} {109}},\ \bibinfo {pages} {18488} (\bibinfo {year}
  {2012})}\BibitemShut {NoStop}%
\bibitem [{\citenamefont {Charlesworth}(2009)}]{Charlesworth2009}%
  \BibitemOpen
  \bibfield  {author} {\bibinfo {author} {\bibfnamefont {B.}~\bibnamefont
  {Charlesworth}},\ }\href@noop {} {\bibfield  {journal} {\bibinfo  {journal}
  {Nature Rev. Genet.}\ }\textbf {\bibinfo {volume} {10}},\ \bibinfo {pages}
  {195} (\bibinfo {year} {2009})}\BibitemShut {NoStop}%
\bibitem [{\citenamefont {Lynch}\ and\ \citenamefont
  {Conery}(2003)}]{Lynch2003}%
  \BibitemOpen
  \bibfield  {author} {\bibinfo {author} {\bibfnamefont {M.}~\bibnamefont
  {Lynch}}\ and\ \bibinfo {author} {\bibfnamefont {J.~S.}\ \bibnamefont
  {Conery}},\ }\href@noop {} {\bibfield  {journal} {\bibinfo  {journal}
  {Science}\ }\textbf {\bibinfo {volume} {302}},\ \bibinfo {pages} {1401}
  (\bibinfo {year} {2003})}\BibitemShut {NoStop}%
\bibitem [{\citenamefont {Lynch}(2005)}]{Lynch2005}%
  \BibitemOpen
  \bibfield  {author} {\bibinfo {author} {\bibfnamefont {M.}~\bibnamefont
  {Lynch}},\ }\href@noop {} {\bibfield  {journal} {\bibinfo  {journal} {Mol.
  Biol. Evol.}\ }\textbf {\bibinfo {volume} {23}},\ \bibinfo {pages} {450}
  (\bibinfo {year} {2005})}\BibitemShut {NoStop}%
\bibitem [{\citenamefont {Koonin}(2016)}]{Koonin2016}%
  \BibitemOpen
  \bibfield  {author} {\bibinfo {author} {\bibfnamefont {E.~V.}\ \bibnamefont
  {Koonin}},\ }\href@noop {} {\bibfield  {journal} {\bibinfo  {journal} {BMC
  Biol.}\ }\textbf {\bibinfo {volume} {14}},\ \bibinfo {pages} {114} (\bibinfo
  {year} {2016})}\BibitemShut {NoStop}%
\bibitem [{\citenamefont {Sela}\ \emph {et~al.}(2016)\citenamefont {Sela},
  \citenamefont {Wolf},\ and\ \citenamefont {Koonin}}]{Sela2016}%
  \BibitemOpen
  \bibfield  {author} {\bibinfo {author} {\bibfnamefont {I.}~\bibnamefont
  {Sela}}, \bibinfo {author} {\bibfnamefont {Y.~I.}\ \bibnamefont {Wolf}}, \
  and\ \bibinfo {author} {\bibfnamefont {E.~V.}\ \bibnamefont {Koonin}},\
  }\href@noop {} {\bibfield  {journal} {\bibinfo  {journal} {Proc. Natl. Acad.
  Sci.}\ }\textbf {\bibinfo {volume} {113}},\ \bibinfo {pages} {11399}
  (\bibinfo {year} {2016})}\BibitemShut {NoStop}%
\bibitem [{\citenamefont {Scott}\ \emph {et~al.}(2010)\citenamefont {Scott},
  \citenamefont {Gunderson}, \citenamefont {Mateescu}, \citenamefont {Zhang},\
  and\ \citenamefont {Hwa}}]{Scott2010}%
  \BibitemOpen
  \bibfield  {author} {\bibinfo {author} {\bibfnamefont {M.}~\bibnamefont
  {Scott}}, \bibinfo {author} {\bibfnamefont {C.~W.}\ \bibnamefont
  {Gunderson}}, \bibinfo {author} {\bibfnamefont {E.~M.}\ \bibnamefont
  {Mateescu}}, \bibinfo {author} {\bibfnamefont {Z.}~\bibnamefont {Zhang}}, \
  and\ \bibinfo {author} {\bibfnamefont {T.}~\bibnamefont {Hwa}},\ }\href@noop
  {} {\bibfield  {journal} {\bibinfo  {journal} {Science}\ }\textbf {\bibinfo
  {volume} {330}},\ \bibinfo {pages} {1099} (\bibinfo {year}
  {2010})}\BibitemShut {NoStop}%
\bibitem [{\citenamefont {Taft}\ \emph {et~al.}(2007)\citenamefont {Taft},
  \citenamefont {Pheasant},\ and\ \citenamefont {Mattick}}]{Taft2007}%
  \BibitemOpen
  \bibfield  {author} {\bibinfo {author} {\bibfnamefont {R.~J.}\ \bibnamefont
  {Taft}}, \bibinfo {author} {\bibfnamefont {M.}~\bibnamefont {Pheasant}}, \
  and\ \bibinfo {author} {\bibfnamefont {J.~S.}\ \bibnamefont {Mattick}},\
  }\href@noop {} {\bibfield  {journal} {\bibinfo  {journal} {Bioessays}\
  }\textbf {\bibinfo {volume} {29}},\ \bibinfo {pages} {288} (\bibinfo {year}
  {2007})}\BibitemShut {NoStop}%
\bibitem [{\citenamefont {Mahmoudabadi}\ \emph {et~al.}(2017)\citenamefont
  {Mahmoudabadi}, \citenamefont {Milo},\ and\ \citenamefont
  {Phillips}}]{Mahmoudabadi2017}%
  \BibitemOpen
  \bibfield  {author} {\bibinfo {author} {\bibfnamefont {G.}~\bibnamefont
  {Mahmoudabadi}}, \bibinfo {author} {\bibfnamefont {R.}~\bibnamefont {Milo}},
  \ and\ \bibinfo {author} {\bibfnamefont {R.}~\bibnamefont {Phillips}},\
  }\href@noop {} {\bibfield  {journal} {\bibinfo  {journal} {Proc. Natl. Acad.
  Sci.}\ }\textbf {\bibinfo {volume} {114}},\ \bibinfo {pages} {E4324}
  (\bibinfo {year} {2017})}\BibitemShut {NoStop}%
\bibitem [{\citenamefont {West}\ \emph {et~al.}(2001)\citenamefont {West},
  \citenamefont {Brown},\ and\ \citenamefont {Enquist}}]{West2001}%
  \BibitemOpen
  \bibfield  {author} {\bibinfo {author} {\bibfnamefont {G.~B.}\ \bibnamefont
  {West}}, \bibinfo {author} {\bibfnamefont {J.~H.}\ \bibnamefont {Brown}}, \
  and\ \bibinfo {author} {\bibfnamefont {B.~J.}\ \bibnamefont {Enquist}},\
  }\href@noop {} {\bibfield  {journal} {\bibinfo  {journal} {Nature}\ }\textbf
  {\bibinfo {volume} {413}},\ \bibinfo {pages} {628} (\bibinfo {year}
  {2001})}\BibitemShut {NoStop}%
\bibitem [{\citenamefont {Hou}\ \emph {et~al.}(2008)\citenamefont {Hou},
  \citenamefont {Zuo}, \citenamefont {Moses}, \citenamefont {Woodruff},
  \citenamefont {Brown},\ and\ \citenamefont {West}}]{Hou2008}%
  \BibitemOpen
  \bibfield  {author} {\bibinfo {author} {\bibfnamefont {C.}~\bibnamefont
  {Hou}}, \bibinfo {author} {\bibfnamefont {W.}~\bibnamefont {Zuo}}, \bibinfo
  {author} {\bibfnamefont {M.~E.}\ \bibnamefont {Moses}}, \bibinfo {author}
  {\bibfnamefont {W.~H.}\ \bibnamefont {Woodruff}}, \bibinfo {author}
  {\bibfnamefont {J.~H.}\ \bibnamefont {Brown}}, \ and\ \bibinfo {author}
  {\bibfnamefont {G.~B.}\ \bibnamefont {West}},\ }\href@noop {} {\bibfield
  {journal} {\bibinfo  {journal} {Science}\ }\textbf {\bibinfo {volume}
  {322}},\ \bibinfo {pages} {736} (\bibinfo {year} {2008})}\BibitemShut
  {NoStop}%
\bibitem [{\citenamefont {Kempes}\ \emph {et~al.}(2012)\citenamefont {Kempes},
  \citenamefont {Dutkiewicz},\ and\ \citenamefont {Follows}}]{Kempes2012}%
  \BibitemOpen
  \bibfield  {author} {\bibinfo {author} {\bibfnamefont {C.~P.}\ \bibnamefont
  {Kempes}}, \bibinfo {author} {\bibfnamefont {S.}~\bibnamefont {Dutkiewicz}},
  \ and\ \bibinfo {author} {\bibfnamefont {M.~J.}\ \bibnamefont {Follows}},\
  }\href@noop {} {\bibfield  {journal} {\bibinfo  {journal} {Proc. Natl. Acad.
  Sci.}\ }\textbf {\bibinfo {volume} {109}},\ \bibinfo {pages} {495} (\bibinfo
  {year} {2012})}\BibitemShut {NoStop}%
\bibitem [{\citenamefont {Pirt}(1965)}]{Pirt1965}%
  \BibitemOpen
  \bibfield  {author} {\bibinfo {author} {\bibfnamefont {S.}~\bibnamefont
  {Pirt}},\ }\href@noop {} {\bibfield  {journal} {\bibinfo  {journal} {Proc. R.
  Soc. Lond. B}\ }\textbf {\bibinfo {volume} {163}},\ \bibinfo {pages} {224}
  (\bibinfo {year} {1965})}\BibitemShut {NoStop}%
\bibitem [{\citenamefont {Kleiber}(1932)}]{Kleiber1932}%
  \BibitemOpen
  \bibfield  {author} {\bibinfo {author} {\bibfnamefont {M.}~\bibnamefont
  {Kleiber}},\ }\href@noop {} {\bibfield  {journal} {\bibinfo  {journal}
  {Hilgardia}\ }\textbf {\bibinfo {volume} {6}},\ \bibinfo {pages} {315}
  (\bibinfo {year} {1932})}\BibitemShut {NoStop}%
\bibitem [{\citenamefont {DeLong}\ \emph {et~al.}(2010)\citenamefont {DeLong},
  \citenamefont {Okie}, \citenamefont {Moses}, \citenamefont {Sibly},\ and\
  \citenamefont {Brown}}]{Delong2010}%
  \BibitemOpen
  \bibfield  {author} {\bibinfo {author} {\bibfnamefont {J.~P.}\ \bibnamefont
  {DeLong}}, \bibinfo {author} {\bibfnamefont {J.~G.}\ \bibnamefont {Okie}},
  \bibinfo {author} {\bibfnamefont {M.~E.}\ \bibnamefont {Moses}}, \bibinfo
  {author} {\bibfnamefont {R.~M.}\ \bibnamefont {Sibly}}, \ and\ \bibinfo
  {author} {\bibfnamefont {J.~H.}\ \bibnamefont {Brown}},\ }\href@noop {}
  {\bibfield  {journal} {\bibinfo  {journal} {Proc. Natl. Acad. Sci.}\ }\textbf
  {\bibinfo {volume} {107}},\ \bibinfo {pages} {12941} (\bibinfo {year}
  {2010})}\BibitemShut {NoStop}%
\bibitem [{\citenamefont {Ballesteros}\ \emph {et~al.}(2018)\citenamefont
  {Ballesteros}, \citenamefont {Martinez}, \citenamefont {Luque}, \citenamefont
  {Lacasa}, \citenamefont {Valor},\ and\ \citenamefont
  {Moya}}]{Ballesteros2018}%
  \BibitemOpen
  \bibfield  {author} {\bibinfo {author} {\bibfnamefont {F.~J.}\ \bibnamefont
  {Ballesteros}}, \bibinfo {author} {\bibfnamefont {V.~J.}\ \bibnamefont
  {Martinez}}, \bibinfo {author} {\bibfnamefont {B.}~\bibnamefont {Luque}},
  \bibinfo {author} {\bibfnamefont {L.}~\bibnamefont {Lacasa}}, \bibinfo
  {author} {\bibfnamefont {E.}~\bibnamefont {Valor}}, \ and\ \bibinfo {author}
  {\bibfnamefont {A.}~\bibnamefont {Moya}},\ }\href@noop {} {\bibfield
  {journal} {\bibinfo  {journal} {Sci. Rep.}\ }\textbf {\bibinfo {volume}
  {8}},\ \bibinfo {pages} {1448} (\bibinfo {year} {2018})}\BibitemShut
  {NoStop}%
\bibitem [{\citenamefont {White}\ and\ \citenamefont
  {Seymour}(2003)}]{White2003}%
  \BibitemOpen
  \bibfield  {author} {\bibinfo {author} {\bibfnamefont {C.~R.}\ \bibnamefont
  {White}}\ and\ \bibinfo {author} {\bibfnamefont {R.~S.}\ \bibnamefont
  {Seymour}},\ }\href@noop {} {\bibfield  {journal} {\bibinfo  {journal} {Proc.
  Natl. Acad. Sci.}\ }\textbf {\bibinfo {volume} {100}},\ \bibinfo {pages}
  {4046} (\bibinfo {year} {2003})}\BibitemShut {NoStop}%
\bibitem [{\citenamefont {Glazier}(2005)}]{Glazier2005}%
  \BibitemOpen
  \bibfield  {author} {\bibinfo {author} {\bibfnamefont {D.~S.}\ \bibnamefont
  {Glazier}},\ }\href@noop {} {\bibfield  {journal} {\bibinfo  {journal} {Biol.
  Rev.}\ }\textbf {\bibinfo {volume} {80}},\ \bibinfo {pages} {611} (\bibinfo
  {year} {2005})}\BibitemShut {NoStop}%
\bibitem [{\citenamefont {Werner}\ \emph {et~al.}(2018)\citenamefont {Werner},
  \citenamefont {Sfakianakis}, \citenamefont {Rendall},\ and\ \citenamefont
  {Griebeler}}]{Werner2018}%
  \BibitemOpen
  \bibfield  {author} {\bibinfo {author} {\bibfnamefont {J.}~\bibnamefont
  {Werner}}, \bibinfo {author} {\bibfnamefont {N.}~\bibnamefont {Sfakianakis}},
  \bibinfo {author} {\bibfnamefont {A.~D.}\ \bibnamefont {Rendall}}, \ and\
  \bibinfo {author} {\bibfnamefont {E.~M.}\ \bibnamefont {Griebeler}},\
  }\href@noop {} {\bibfield  {journal} {\bibinfo  {journal} {J. Theor. Biol.}\
  }\textbf {\bibinfo {volume} {444}},\ \bibinfo {pages} {83} (\bibinfo {year}
  {2018})}\BibitemShut {NoStop}%
\bibitem [{\citenamefont {Dietz}\ and\ \citenamefont
  {Drent}(1997)}]{Dietz1997}%
  \BibitemOpen
  \bibfield  {author} {\bibinfo {author} {\bibfnamefont {M.~W.}\ \bibnamefont
  {Dietz}}\ and\ \bibinfo {author} {\bibfnamefont {R.~H.}\ \bibnamefont
  {Drent}},\ }\href@noop {} {\bibfield  {journal} {\bibinfo  {journal}
  {Physiol. Zool.}\ }\textbf {\bibinfo {volume} {70}},\ \bibinfo {pages} {493}
  (\bibinfo {year} {1997})}\BibitemShut {NoStop}%
\bibitem [{\citenamefont {May}(1976)}]{May1976}%
  \BibitemOpen
  \bibfield  {author} {\bibinfo {author} {\bibfnamefont {R.~M.}\ \bibnamefont
  {May}},\ }\href@noop {} {\bibfield  {journal} {\bibinfo  {journal} {Am.
  Nat.}\ }\textbf {\bibinfo {volume} {110}},\ \bibinfo {pages} {496} (\bibinfo
  {year} {1976})}\BibitemShut {NoStop}%
\bibitem [{\citenamefont {Savage}\ \emph {et~al.}(2004)\citenamefont {Savage},
  \citenamefont {Gillooly}, \citenamefont {Brown}, \citenamefont {West},\ and\
  \citenamefont {Charnov}}]{Savage2004}%
  \BibitemOpen
  \bibfield  {author} {\bibinfo {author} {\bibfnamefont {V.~M.}\ \bibnamefont
  {Savage}}, \bibinfo {author} {\bibfnamefont {J.~F.}\ \bibnamefont
  {Gillooly}}, \bibinfo {author} {\bibfnamefont {J.~H.}\ \bibnamefont {Brown}},
  \bibinfo {author} {\bibfnamefont {G.~B.}\ \bibnamefont {West}}, \ and\
  \bibinfo {author} {\bibfnamefont {E.~L.}\ \bibnamefont {Charnov}},\
  }\href@noop {} {\bibfield  {journal} {\bibinfo  {journal} {Am. Nat.}\
  }\textbf {\bibinfo {volume} {163}},\ \bibinfo {pages} {429} (\bibinfo {year}
  {2004})}\BibitemShut {NoStop}%
\bibitem [{\citenamefont {Moses}\ \emph {et~al.}(2008)\citenamefont {Moses},
  \citenamefont {Hou}, \citenamefont {Woodruff}, \citenamefont {West},
  \citenamefont {Nekola}, \citenamefont {Zuo},\ and\ \citenamefont
  {Brown}}]{Moses2008}%
  \BibitemOpen
  \bibfield  {author} {\bibinfo {author} {\bibfnamefont {M.~E.}\ \bibnamefont
  {Moses}}, \bibinfo {author} {\bibfnamefont {C.}~\bibnamefont {Hou}}, \bibinfo
  {author} {\bibfnamefont {W.~H.}\ \bibnamefont {Woodruff}}, \bibinfo {author}
  {\bibfnamefont {G.~B.}\ \bibnamefont {West}}, \bibinfo {author}
  {\bibfnamefont {J.~C.}\ \bibnamefont {Nekola}}, \bibinfo {author}
  {\bibfnamefont {W.}~\bibnamefont {Zuo}}, \ and\ \bibinfo {author}
  {\bibfnamefont {J.~H.}\ \bibnamefont {Brown}},\ }\href@noop {} {\bibfield
  {journal} {\bibinfo  {journal} {Am. Nat.}\ }\textbf {\bibinfo {volume}
  {171}},\ \bibinfo {pages} {632} (\bibinfo {year} {2008})}\BibitemShut
  {NoStop}%
\bibitem [{\citenamefont {Maitra}\ and\ \citenamefont
  {Dill}(2015)}]{Maitra2015}%
  \BibitemOpen
  \bibfield  {author} {\bibinfo {author} {\bibfnamefont {A.}~\bibnamefont
  {Maitra}}\ and\ \bibinfo {author} {\bibfnamefont {K.~A.}\ \bibnamefont
  {Dill}},\ }\href@noop {} {\bibfield  {journal} {\bibinfo  {journal} {Proc.
  Natl. Acad. Sci.}\ }\textbf {\bibinfo {volume} {112}},\ \bibinfo {pages}
  {406} (\bibinfo {year} {2015})}\BibitemShut {NoStop}%
\bibitem [{\citenamefont {Berkhout}\ \emph {et~al.}(2013)\citenamefont
  {Berkhout}, \citenamefont {Bosdriesz}, \citenamefont {Nikerel}, \citenamefont
  {Molenaar}, \citenamefont {de~Ridder}, \citenamefont {Teusink},\ and\
  \citenamefont {Bruggeman}}]{Berkhout2013}%
  \BibitemOpen
  \bibfield  {author} {\bibinfo {author} {\bibfnamefont {J.}~\bibnamefont
  {Berkhout}}, \bibinfo {author} {\bibfnamefont {E.}~\bibnamefont {Bosdriesz}},
  \bibinfo {author} {\bibfnamefont {E.}~\bibnamefont {Nikerel}}, \bibinfo
  {author} {\bibfnamefont {D.}~\bibnamefont {Molenaar}}, \bibinfo {author}
  {\bibfnamefont {D.}~\bibnamefont {de~Ridder}}, \bibinfo {author}
  {\bibfnamefont {B.}~\bibnamefont {Teusink}}, \ and\ \bibinfo {author}
  {\bibfnamefont {F.~J.}\ \bibnamefont {Bruggeman}},\ }\href@noop {} {\bibfield
   {journal} {\bibinfo  {journal} {Genetics}\ }\textbf {\bibinfo {volume}
  {194}},\ \bibinfo {pages} {505} (\bibinfo {year} {2013})}\BibitemShut
  {NoStop}%
\bibitem [{\citenamefont {Price}\ and\ \citenamefont
  {Arkin}(2016)}]{Price2016}%
  \BibitemOpen
  \bibfield  {author} {\bibinfo {author} {\bibfnamefont {M.~N.}\ \bibnamefont
  {Price}}\ and\ \bibinfo {author} {\bibfnamefont {A.~P.}\ \bibnamefont
  {Arkin}},\ }\href@noop {} {\bibfield  {journal} {\bibinfo  {journal} {Genome
  Biol. Evol.}\ }\textbf {\bibinfo {volume} {8}},\ \bibinfo {pages} {1917}
  (\bibinfo {year} {2016})}\BibitemShut {NoStop}%
\bibitem [{\citenamefont {Grilli}\ \emph {et~al.}(2017)\citenamefont {Grilli},
  \citenamefont {Osella}, \citenamefont {Kennard},\ and\ \citenamefont
  {Lagomarsino}}]{Grilli2017}%
  \BibitemOpen
  \bibfield  {author} {\bibinfo {author} {\bibfnamefont {J.}~\bibnamefont
  {Grilli}}, \bibinfo {author} {\bibfnamefont {M.}~\bibnamefont {Osella}},
  \bibinfo {author} {\bibfnamefont {A.~S.}\ \bibnamefont {Kennard}}, \ and\
  \bibinfo {author} {\bibfnamefont {M.~C.}\ \bibnamefont {Lagomarsino}},\
  }\href@noop {} {\bibfield  {journal} {\bibinfo  {journal} {Physical Review
  E}\ }\textbf {\bibinfo {volume} {95}},\ \bibinfo {pages} {032411} (\bibinfo
  {year} {2017})}\BibitemShut {NoStop}%
\bibitem [{\citenamefont {Pfleger}\ \emph {et~al.}(2015)\citenamefont
  {Pfleger}, \citenamefont {He},\ and\ \citenamefont
  {Abdellatif}}]{Pfleger2015}%
  \BibitemOpen
  \bibfield  {author} {\bibinfo {author} {\bibfnamefont {J.}~\bibnamefont
  {Pfleger}}, \bibinfo {author} {\bibfnamefont {M.}~\bibnamefont {He}}, \ and\
  \bibinfo {author} {\bibfnamefont {M.}~\bibnamefont {Abdellatif}},\
  }\href@noop {} {\bibfield  {journal} {\bibinfo  {journal} {Cell Death Dis.}\
  }\textbf {\bibinfo {volume} {6}},\ \bibinfo {pages} {e1835} (\bibinfo {year}
  {2015})}\BibitemShut {NoStop}%
\bibitem [{\citenamefont {Choi}\ \emph {et~al.}(2009)\citenamefont {Choi},
  \citenamefont {Gerencser},\ and\ \citenamefont {Nicholls}}]{Choi2009}%
  \BibitemOpen
  \bibfield  {author} {\bibinfo {author} {\bibfnamefont {S.~W.}\ \bibnamefont
  {Choi}}, \bibinfo {author} {\bibfnamefont {A.~A.}\ \bibnamefont {Gerencser}},
  \ and\ \bibinfo {author} {\bibfnamefont {D.~G.}\ \bibnamefont {Nicholls}},\
  }\href@noop {} {\bibfield  {journal} {\bibinfo  {journal} {J. Neurochem.}\
  }\textbf {\bibinfo {volume} {109}},\ \bibinfo {pages} {1179} (\bibinfo {year}
  {2009})}\BibitemShut {NoStop}%
\bibitem [{\citenamefont {Yaginuma}\ \emph {et~al.}(2014)\citenamefont
  {Yaginuma}, \citenamefont {Kawai}, \citenamefont {Tabata}, \citenamefont
  {Tomiyama}, \citenamefont {Kakizuka}, \citenamefont {Komatsuzaki},
  \citenamefont {Noji},\ and\ \citenamefont {Imamura}}]{Yaginuma2014}%
  \BibitemOpen
  \bibfield  {author} {\bibinfo {author} {\bibfnamefont {H.}~\bibnamefont
  {Yaginuma}}, \bibinfo {author} {\bibfnamefont {S.}~\bibnamefont {Kawai}},
  \bibinfo {author} {\bibfnamefont {K.~V.}\ \bibnamefont {Tabata}}, \bibinfo
  {author} {\bibfnamefont {K.}~\bibnamefont {Tomiyama}}, \bibinfo {author}
  {\bibfnamefont {A.}~\bibnamefont {Kakizuka}}, \bibinfo {author}
  {\bibfnamefont {T.}~\bibnamefont {Komatsuzaki}}, \bibinfo {author}
  {\bibfnamefont {H.}~\bibnamefont {Noji}}, \ and\ \bibinfo {author}
  {\bibfnamefont {H.}~\bibnamefont {Imamura}},\ }\href@noop {} {\bibfield
  {journal} {\bibinfo  {journal} {Sci. Rep.}\ }\textbf {\bibinfo {volume}
  {4}},\ \bibinfo {pages} {6522} (\bibinfo {year} {2014})}\BibitemShut
  {NoStop}%
\bibitem [{\citenamefont {Nickens}\ \emph {et~al.}(2013)\citenamefont
  {Nickens}, \citenamefont {Wikstrom}, \citenamefont {Shirihai}, \citenamefont
  {Patierno},\ and\ \citenamefont {Ceryak}}]{Nickens2013}%
  \BibitemOpen
  \bibfield  {author} {\bibinfo {author} {\bibfnamefont {K.~P.}\ \bibnamefont
  {Nickens}}, \bibinfo {author} {\bibfnamefont {J.~D.}\ \bibnamefont
  {Wikstrom}}, \bibinfo {author} {\bibfnamefont {O.~S.}\ \bibnamefont
  {Shirihai}}, \bibinfo {author} {\bibfnamefont {S.~R.}\ \bibnamefont
  {Patierno}}, \ and\ \bibinfo {author} {\bibfnamefont {S.}~\bibnamefont
  {Ceryak}},\ }\href@noop {} {\bibfield  {journal} {\bibinfo  {journal}
  {Mitochondrion}\ }\textbf {\bibinfo {volume} {13}},\ \bibinfo {pages} {662}
  (\bibinfo {year} {2013})}\BibitemShut {NoStop}%
\bibitem [{\citenamefont {Brand}\ and\ \citenamefont
  {Nicholls}(2011)}]{Brand2011}%
  \BibitemOpen
  \bibfield  {author} {\bibinfo {author} {\bibfnamefont {M.~D.}\ \bibnamefont
  {Brand}}\ and\ \bibinfo {author} {\bibfnamefont {D.~G.}\ \bibnamefont
  {Nicholls}},\ }\href@noop {} {\bibfield  {journal} {\bibinfo  {journal}
  {Biochem. J.}\ }\textbf {\bibinfo {volume} {435}},\ \bibinfo {pages} {297}
  (\bibinfo {year} {2011})}\BibitemShut {NoStop}%
\bibitem [{\citenamefont {Sauls}\ \emph {et~al.}(2016)\citenamefont {Sauls},
  \citenamefont {Li},\ and\ \citenamefont {Jun}}]{Sauls2016}%
  \BibitemOpen
  \bibfield  {author} {\bibinfo {author} {\bibfnamefont {J.~T.}\ \bibnamefont
  {Sauls}}, \bibinfo {author} {\bibfnamefont {D.}~\bibnamefont {Li}}, \ and\
  \bibinfo {author} {\bibfnamefont {S.}~\bibnamefont {Jun}},\ }\href@noop {}
  {\bibfield  {journal} {\bibinfo  {journal} {Curr. Opin. Cell Biol.}\ }\textbf
  {\bibinfo {volume} {38}},\ \bibinfo {pages} {38} (\bibinfo {year}
  {2016})}\BibitemShut {NoStop}%
\bibitem [{\citenamefont {Facchetti}\ \emph {et~al.}(2017)\citenamefont
  {Facchetti}, \citenamefont {Chang},\ and\ \citenamefont
  {Howard}}]{Facchetti2017}%
  \BibitemOpen
  \bibfield  {author} {\bibinfo {author} {\bibfnamefont {G.}~\bibnamefont
  {Facchetti}}, \bibinfo {author} {\bibfnamefont {F.}~\bibnamefont {Chang}}, \
  and\ \bibinfo {author} {\bibfnamefont {M.}~\bibnamefont {Howard}},\
  }\href@noop {} {\bibfield  {journal} {\bibinfo  {journal} {Curr. Opin. Sys.
  Biol.}\ }\textbf {\bibinfo {volume} {5}},\ \bibinfo {pages} {86} (\bibinfo
  {year} {2017})}\BibitemShut {NoStop}%
\bibitem [{\citenamefont {Martinez}\ \emph {et~al.}(2008)\citenamefont
  {Martinez}, \citenamefont {Fajardo}, \citenamefont {Garmendia}, \citenamefont
  {Hernandez}, \citenamefont {Linares}, \citenamefont {Mart{\'\i}nez-Solano},\
  and\ \citenamefont {S{\'a}nchez}}]{Martinez2008}%
  \BibitemOpen
  \bibfield  {author} {\bibinfo {author} {\bibfnamefont {J.~L.}\ \bibnamefont
  {Martinez}}, \bibinfo {author} {\bibfnamefont {A.}~\bibnamefont {Fajardo}},
  \bibinfo {author} {\bibfnamefont {L.}~\bibnamefont {Garmendia}}, \bibinfo
  {author} {\bibfnamefont {A.}~\bibnamefont {Hernandez}}, \bibinfo {author}
  {\bibfnamefont {J.~F.}\ \bibnamefont {Linares}}, \bibinfo {author}
  {\bibfnamefont {L.}~\bibnamefont {Mart{\'\i}nez-Solano}}, \ and\ \bibinfo
  {author} {\bibfnamefont {M.~B.}\ \bibnamefont {S{\'a}nchez}},\ }\href@noop {}
  {\bibfield  {journal} {\bibinfo  {journal} {FEMS Microbiol. Rev.}\ }\textbf
  {\bibinfo {volume} {33}},\ \bibinfo {pages} {44} (\bibinfo {year}
  {2008})}\BibitemShut {NoStop}%
\bibitem [{\citenamefont {Blanco}\ \emph {et~al.}(2016)\citenamefont {Blanco},
  \citenamefont {Hernando-Amado}, \citenamefont {Reales-Calderon},
  \citenamefont {Corona}, \citenamefont {Lira}, \citenamefont {Alcalde-Rico},
  \citenamefont {Bernardini}, \citenamefont {Sanchez},\ and\ \citenamefont
  {Martinez}}]{Blanco2016}%
  \BibitemOpen
  \bibfield  {author} {\bibinfo {author} {\bibfnamefont {P.}~\bibnamefont
  {Blanco}}, \bibinfo {author} {\bibfnamefont {S.}~\bibnamefont
  {Hernando-Amado}}, \bibinfo {author} {\bibfnamefont {J.}~\bibnamefont
  {Reales-Calderon}}, \bibinfo {author} {\bibfnamefont {F.}~\bibnamefont
  {Corona}}, \bibinfo {author} {\bibfnamefont {F.}~\bibnamefont {Lira}},
  \bibinfo {author} {\bibfnamefont {M.}~\bibnamefont {Alcalde-Rico}}, \bibinfo
  {author} {\bibfnamefont {A.}~\bibnamefont {Bernardini}}, \bibinfo {author}
  {\bibfnamefont {M.}~\bibnamefont {Sanchez}}, \ and\ \bibinfo {author}
  {\bibfnamefont {J.}~\bibnamefont {Martinez}},\ }\href@noop {} {\bibfield
  {journal} {\bibinfo  {journal} {Microorganisms}\ }\textbf {\bibinfo {volume}
  {4}},\ \bibinfo {pages} {14} (\bibinfo {year} {2016})}\BibitemShut {NoStop}%
\bibitem [{\citenamefont {Ilker}\ and\ \citenamefont
  {Hinczewski}(2019)}]{Ilker2019}%
  \BibitemOpen
  \bibfield  {author} {\bibinfo {author} {\bibfnamefont {E.}~\bibnamefont
  {Ilker}}\ and\ \bibinfo {author} {\bibfnamefont {M.}~\bibnamefont
  {Hinczewski}},\ }\href@noop {} {}\bibinfo {howpublished} {in preparation}
  (\bibinfo {year} {2019})\BibitemShut {NoStop}%
\bibitem [{\citenamefont {Zampieri}\ \emph {et~al.}(2017)\citenamefont
  {Zampieri}, \citenamefont {Enke}, \citenamefont {Chubukov}, \citenamefont
  {Ricci}, \citenamefont {Piddock},\ and\ \citenamefont
  {Sauer}}]{Zampieri2017}%
  \BibitemOpen
  \bibfield  {author} {\bibinfo {author} {\bibfnamefont {M.}~\bibnamefont
  {Zampieri}}, \bibinfo {author} {\bibfnamefont {T.}~\bibnamefont {Enke}},
  \bibinfo {author} {\bibfnamefont {V.}~\bibnamefont {Chubukov}}, \bibinfo
  {author} {\bibfnamefont {V.}~\bibnamefont {Ricci}}, \bibinfo {author}
  {\bibfnamefont {L.}~\bibnamefont {Piddock}}, \ and\ \bibinfo {author}
  {\bibfnamefont {U.}~\bibnamefont {Sauer}},\ }\href@noop {} {\bibfield
  {journal} {\bibinfo  {journal} {Mol. Sys. Biol.}\ }\textbf {\bibinfo {volume}
  {13}},\ \bibinfo {pages} {917} (\bibinfo {year} {2017})}\BibitemShut
  {NoStop}%
\bibitem [{\citenamefont {Dahlberg}\ and\ \citenamefont
  {Chao}(2003)}]{Dahlberg2003}%
  \BibitemOpen
  \bibfield  {author} {\bibinfo {author} {\bibfnamefont {C.}~\bibnamefont
  {Dahlberg}}\ and\ \bibinfo {author} {\bibfnamefont {L.}~\bibnamefont
  {Chao}},\ }\href@noop {} {\bibfield  {journal} {\bibinfo  {journal}
  {Genetics}\ }\textbf {\bibinfo {volume} {165}},\ \bibinfo {pages} {1641}
  (\bibinfo {year} {2003})}\BibitemShut {NoStop}%
\bibitem [{\citenamefont {Melnyk}\ \emph {et~al.}(2015)\citenamefont {Melnyk},
  \citenamefont {Wong},\ and\ \citenamefont {Kassen}}]{Melnyk2015}%
  \BibitemOpen
  \bibfield  {author} {\bibinfo {author} {\bibfnamefont {A.~H.}\ \bibnamefont
  {Melnyk}}, \bibinfo {author} {\bibfnamefont {A.}~\bibnamefont {Wong}}, \ and\
  \bibinfo {author} {\bibfnamefont {R.}~\bibnamefont {Kassen}},\ }\href@noop {}
  {\bibfield  {journal} {\bibinfo  {journal} {Evol. Appl.}\ }\textbf {\bibinfo
  {volume} {8}},\ \bibinfo {pages} {273} (\bibinfo {year} {2015})}\BibitemShut
  {NoStop}%
\bibitem [{\citenamefont {Pawar}\ \emph {et~al.}(2012)\citenamefont {Pawar},
  \citenamefont {Dell},\ and\ \citenamefont {Savage}}]{Pawar2012}%
  \BibitemOpen
  \bibfield  {author} {\bibinfo {author} {\bibfnamefont {S.}~\bibnamefont
  {Pawar}}, \bibinfo {author} {\bibfnamefont {A.~I.}\ \bibnamefont {Dell}}, \
  and\ \bibinfo {author} {\bibfnamefont {V.~M.}\ \bibnamefont {Savage}},\
  }\href@noop {} {\bibfield  {journal} {\bibinfo  {journal} {Nature}\ }\textbf
  {\bibinfo {volume} {486}},\ \bibinfo {pages} {485} (\bibinfo {year}
  {2012})}\BibitemShut {NoStop}%
\bibitem [{\citenamefont {Karr}\ \emph {et~al.}(2012)\citenamefont {Karr},
  \citenamefont {Sanghvi}, \citenamefont {Macklin}, \citenamefont {Gutschow},
  \citenamefont {Jacobs}, \citenamefont {Bolival~Jr}, \citenamefont
  {Assad-Garcia}, \citenamefont {Glass},\ and\ \citenamefont
  {Covert}}]{Karr2012}%
  \BibitemOpen
  \bibfield  {author} {\bibinfo {author} {\bibfnamefont {J.~R.}\ \bibnamefont
  {Karr}}, \bibinfo {author} {\bibfnamefont {J.~C.}\ \bibnamefont {Sanghvi}},
  \bibinfo {author} {\bibfnamefont {D.~N.}\ \bibnamefont {Macklin}}, \bibinfo
  {author} {\bibfnamefont {M.~V.}\ \bibnamefont {Gutschow}}, \bibinfo {author}
  {\bibfnamefont {J.~M.}\ \bibnamefont {Jacobs}}, \bibinfo {author}
  {\bibfnamefont {B.}~\bibnamefont {Bolival~Jr}}, \bibinfo {author}
  {\bibfnamefont {N.}~\bibnamefont {Assad-Garcia}}, \bibinfo {author}
  {\bibfnamefont {J.~I.}\ \bibnamefont {Glass}}, \ and\ \bibinfo {author}
  {\bibfnamefont {M.~W.}\ \bibnamefont {Covert}},\ }\href@noop {} {\bibfield
  {journal} {\bibinfo  {journal} {Cell}\ }\textbf {\bibinfo {volume} {150}},\
  \bibinfo {pages} {389} (\bibinfo {year} {2012})}\BibitemShut {NoStop}%
\bibitem [{\citenamefont {Marantan}\ and\ \citenamefont
  {Amir}(2016)}]{Marantan2016}%
  \BibitemOpen
  \bibfield  {author} {\bibinfo {author} {\bibfnamefont {A.}~\bibnamefont
  {Marantan}}\ and\ \bibinfo {author} {\bibfnamefont {A.}~\bibnamefont
  {Amir}},\ }\href@noop {} {\bibfield  {journal} {\bibinfo  {journal} {Phys.
  Rev. E}\ }\textbf {\bibinfo {volume} {94}},\ \bibinfo {pages} {012405}
  (\bibinfo {year} {2016})}\BibitemShut {NoStop}%
\bibitem [{\citenamefont {Sibly}\ and\ \citenamefont
  {Brown}(2009)}]{Sibly2009}%
  \BibitemOpen
  \bibfield  {author} {\bibinfo {author} {\bibfnamefont {R.~M.}\ \bibnamefont
  {Sibly}}\ and\ \bibinfo {author} {\bibfnamefont {J.~H.}\ \bibnamefont
  {Brown}},\ }\href@noop {} {\bibfield  {journal} {\bibinfo  {journal} {Am.
  Nat.}\ }\textbf {\bibinfo {volume} {173}},\ \bibinfo {pages} {E185} (\bibinfo
  {year} {2009})}\BibitemShut {NoStop}%
\bibitem [{\citenamefont {Courchamp}\ \emph {et~al.}(2008)\citenamefont
  {Courchamp}, \citenamefont {Berec},\ and\ \citenamefont
  {Gascoigne}}]{Courchamp2008}%
  \BibitemOpen
  \bibfield  {author} {\bibinfo {author} {\bibfnamefont {F.}~\bibnamefont
  {Courchamp}}, \bibinfo {author} {\bibfnamefont {L.}~\bibnamefont {Berec}}, \
  and\ \bibinfo {author} {\bibfnamefont {J.}~\bibnamefont {Gascoigne}},\
  }\href@noop {} {\emph {\bibinfo {title} {Allee effects in ecology and
  conservation}}}\ (\bibinfo  {publisher} {Oxford Univ. Press},\ \bibinfo
  {year} {2008})\BibitemShut {NoStop}%
\bibitem [{\citenamefont {Smith}()}]{Smith}%
  \BibitemOpen
  \bibfield  {author} {\bibinfo {author} {\bibfnamefont {H.~L.}\ \bibnamefont
  {Smith}},\ }\href@noop {} {\enquote {\bibinfo {title} {The
  {Rosenzweig}-{Macarthur} predator-prey model},}\ }\bibinfo {howpublished}
  {\url{https://math.la.asu.edu/~halsmith/Rosenzweig.pdf}}\BibitemShut
  {NoStop}%
\bibitem [{\citenamefont {Holling}(1959)}]{Holling1959}%
  \BibitemOpen
  \bibfield  {author} {\bibinfo {author} {\bibfnamefont {C.~S.}\ \bibnamefont
  {Holling}},\ }\href@noop {} {\bibfield  {journal} {\bibinfo  {journal} {Can.
  Entomol.}\ }\textbf {\bibinfo {volume} {91}},\ \bibinfo {pages} {385}
  (\bibinfo {year} {1959})}\BibitemShut {NoStop}%
\bibitem [{\citenamefont {Turchin}(2003)}]{Turchin2003}%
  \BibitemOpen
  \bibfield  {author} {\bibinfo {author} {\bibfnamefont {P.}~\bibnamefont
  {Turchin}},\ }\href@noop {} {\emph {\bibinfo {title} {Complex population
  dynamics: a theoretical/empirical synthesis}}}\ (\bibinfo  {publisher}
  {Princeton Univ. Press},\ \bibinfo {year} {2003})\BibitemShut {NoStop}%
\bibitem [{\citenamefont {Monod}(1949)}]{Monod1949}%
  \BibitemOpen
  \bibfield  {author} {\bibinfo {author} {\bibfnamefont {J.}~\bibnamefont
  {Monod}},\ }\href@noop {} {\bibfield  {journal} {\bibinfo  {journal} {Annu.
  Rev. in Microbiol.}\ }\textbf {\bibinfo {volume} {3}},\ \bibinfo {pages}
  {371} (\bibinfo {year} {1949})}\BibitemShut {NoStop}%
\bibitem [{\citenamefont {Rosenzweig}\ and\ \citenamefont
  {MacArthur}(1963)}]{Rosenzweig1963}%
  \BibitemOpen
  \bibfield  {author} {\bibinfo {author} {\bibfnamefont {M.~L.}\ \bibnamefont
  {Rosenzweig}}\ and\ \bibinfo {author} {\bibfnamefont {R.~H.}\ \bibnamefont
  {MacArthur}},\ }\href@noop {} {\bibfield  {journal} {\bibinfo  {journal} {Am.
  Nat.}\ }\textbf {\bibinfo {volume} {97}},\ \bibinfo {pages} {209} (\bibinfo
  {year} {1963})}\BibitemShut {NoStop}%
\bibitem [{\citenamefont {Novick}\ and\ \citenamefont
  {Szilard}(1950{\natexlab{a}})}]{Novick1950a}%
  \BibitemOpen
  \bibfield  {author} {\bibinfo {author} {\bibfnamefont {A.}~\bibnamefont
  {Novick}}\ and\ \bibinfo {author} {\bibfnamefont {L.}~\bibnamefont
  {Szilard}},\ }\href@noop {} {\bibfield  {journal} {\bibinfo  {journal}
  {Science}\ }\textbf {\bibinfo {volume} {112}},\ \bibinfo {pages} {715}
  (\bibinfo {year} {1950}{\natexlab{a}})}\BibitemShut {NoStop}%
\bibitem [{\citenamefont {Novick}\ and\ \citenamefont
  {Szilard}(1950{\natexlab{b}})}]{Novick1950b}%
  \BibitemOpen
  \bibfield  {author} {\bibinfo {author} {\bibfnamefont {A.}~\bibnamefont
  {Novick}}\ and\ \bibinfo {author} {\bibfnamefont {L.}~\bibnamefont
  {Szilard}},\ }\href@noop {} {\bibfield  {journal} {\bibinfo  {journal} {Proc.
  Natl. Acad. Sci.}\ }\textbf {\bibinfo {volume} {36}},\ \bibinfo {pages} {708}
  (\bibinfo {year} {1950}{\natexlab{b}})}\BibitemShut {NoStop}%
\bibitem [{\citenamefont {Gresham}\ and\ \citenamefont
  {Hong}(2014)}]{Gresham2014}%
  \BibitemOpen
  \bibfield  {author} {\bibinfo {author} {\bibfnamefont {D.}~\bibnamefont
  {Gresham}}\ and\ \bibinfo {author} {\bibfnamefont {J.}~\bibnamefont {Hong}},\
  }\href@noop {} {\bibfield  {journal} {\bibinfo  {journal} {FEMS Microbiol.
  Rev.}\ }\textbf {\bibinfo {volume} {39}},\ \bibinfo {pages} {2} (\bibinfo
  {year} {2014})}\BibitemShut {NoStop}%
\bibitem [{\citenamefont {De~Magalh{\~a}es}\ \emph {et~al.}(2005)\citenamefont
  {De~Magalh{\~a}es}, \citenamefont {Costa},\ and\ \citenamefont
  {Toussaint}}]{HAGR2005}%
  \BibitemOpen
  \bibfield  {author} {\bibinfo {author} {\bibfnamefont {J.~P.}\ \bibnamefont
  {De~Magalh{\~a}es}}, \bibinfo {author} {\bibfnamefont {J.}~\bibnamefont
  {Costa}}, \ and\ \bibinfo {author} {\bibfnamefont {O.}~\bibnamefont
  {Toussaint}},\ }\href@noop {} {\bibfield  {journal} {\bibinfo  {journal}
  {Nucleic Acids Res.}\ }\textbf {\bibinfo {volume} {33}},\ \bibinfo {pages}
  {D537} (\bibinfo {year} {2005})}\BibitemShut {NoStop}%
\bibitem [{\citenamefont {Sears}\ \emph {et~al.}(2012)\citenamefont {Sears},
  \citenamefont {Kerkhoff}, \citenamefont {Messerman},\ and\ \citenamefont
  {Itagaki}}]{Sears2012}%
  \BibitemOpen
  \bibfield  {author} {\bibinfo {author} {\bibfnamefont {K.~E.}\ \bibnamefont
  {Sears}}, \bibinfo {author} {\bibfnamefont {A.~J.}\ \bibnamefont {Kerkhoff}},
  \bibinfo {author} {\bibfnamefont {A.}~\bibnamefont {Messerman}}, \ and\
  \bibinfo {author} {\bibfnamefont {H.}~\bibnamefont {Itagaki}},\ }\href@noop
  {} {\bibfield  {journal} {\bibinfo  {journal} {Physiol. Biochem. Zool.}\
  }\textbf {\bibinfo {volume} {85}},\ \bibinfo {pages} {159} (\bibinfo {year}
  {2012})}\BibitemShut {NoStop}%
\bibitem [{\citenamefont {Callier}\ and\ \citenamefont
  {Nijhout}(2012)}]{Callier2012}%
  \BibitemOpen
  \bibfield  {author} {\bibinfo {author} {\bibfnamefont {V.}~\bibnamefont
  {Callier}}\ and\ \bibinfo {author} {\bibfnamefont {H.~F.}\ \bibnamefont
  {Nijhout}},\ }\href@noop {} {\bibfield  {journal} {\bibinfo  {journal} {PLoS
  ONE}\ }\textbf {\bibinfo {volume} {7}},\ \bibinfo {pages} {e45455} (\bibinfo
  {year} {2012})}\BibitemShut {NoStop}%
\bibitem [{\citenamefont {Glazier}\ \emph {et~al.}(2015)\citenamefont
  {Glazier}, \citenamefont {Hirst},\ and\ \citenamefont
  {Atkinson}}]{Glazier2018}%
  \BibitemOpen
  \bibfield  {author} {\bibinfo {author} {\bibfnamefont {D.~S.}\ \bibnamefont
  {Glazier}}, \bibinfo {author} {\bibfnamefont {A.~G.}\ \bibnamefont {Hirst}},
  \ and\ \bibinfo {author} {\bibfnamefont {D.}~\bibnamefont {Atkinson}},\
  }\href@noop {} {\bibfield  {journal} {\bibinfo  {journal} {Proc. R. Soc. B}\
  }\textbf {\bibinfo {volume} {282}},\ \bibinfo {pages} {20142302} (\bibinfo
  {year} {2015})}\BibitemShut {NoStop}%
\bibitem [{\citenamefont {Farlow}\ \emph {et~al.}(2015)\citenamefont {Farlow},
  \citenamefont {Long}, \citenamefont {Arnoux}, \citenamefont {Sung},
  \citenamefont {Doak}, \citenamefont {Nordborg},\ and\ \citenamefont
  {Lynch}}]{Farlow2015}%
  \BibitemOpen
  \bibfield  {author} {\bibinfo {author} {\bibfnamefont {A.}~\bibnamefont
  {Farlow}}, \bibinfo {author} {\bibfnamefont {H.}~\bibnamefont {Long}},
  \bibinfo {author} {\bibfnamefont {S.}~\bibnamefont {Arnoux}}, \bibinfo
  {author} {\bibfnamefont {W.}~\bibnamefont {Sung}}, \bibinfo {author}
  {\bibfnamefont {T.~G.}\ \bibnamefont {Doak}}, \bibinfo {author}
  {\bibfnamefont {M.}~\bibnamefont {Nordborg}}, \ and\ \bibinfo {author}
  {\bibfnamefont {M.}~\bibnamefont {Lynch}},\ }\href@noop {} {\bibfield
  {journal} {\bibinfo  {journal} {Genetics}\ }\textbf {\bibinfo {volume}
  {201}},\ \bibinfo {pages} {737} (\bibinfo {year} {2015})}\BibitemShut
  {NoStop}%
\bibitem [{\citenamefont {Charlesworth}\ and\ \citenamefont
  {Eyre-Walker}(2006)}]{Charlesworth2006}%
  \BibitemOpen
  \bibfield  {author} {\bibinfo {author} {\bibfnamefont {J.}~\bibnamefont
  {Charlesworth}}\ and\ \bibinfo {author} {\bibfnamefont {A.}~\bibnamefont
  {Eyre-Walker}},\ }\href@noop {} {\bibfield  {journal} {\bibinfo  {journal}
  {Mol. Biol. Evol.}\ }\textbf {\bibinfo {volume} {23}},\ \bibinfo {pages}
  {1348} (\bibinfo {year} {2006})}\BibitemShut {NoStop}%
\bibitem [{\citenamefont {Chang}(2017)}]{Chang2017}%
  \BibitemOpen
  \bibfield  {author} {\bibinfo {author} {\bibfnamefont {F.}~\bibnamefont
  {Chang}},\ }\href@noop {} {\bibfield  {journal} {\bibinfo  {journal} {Mol.
  Biol. Cell}\ }\textbf {\bibinfo {volume} {28}},\ \bibinfo {pages} {1819}
  (\bibinfo {year} {2017})}\BibitemShut {NoStop}%
\bibitem [{\citenamefont {Raghavan}\ \emph {et~al.}(2011)\citenamefont
  {Raghavan}, \citenamefont {Groisman},\ and\ \citenamefont
  {Ochman}}]{Raghavan2011}%
  \BibitemOpen
  \bibfield  {author} {\bibinfo {author} {\bibfnamefont {R.}~\bibnamefont
  {Raghavan}}, \bibinfo {author} {\bibfnamefont {E.~A.}\ \bibnamefont
  {Groisman}}, \ and\ \bibinfo {author} {\bibfnamefont {H.}~\bibnamefont
  {Ochman}},\ }\href@noop {} {\bibfield  {journal} {\bibinfo  {journal} {Genome
  Res.}\ }\textbf {\bibinfo {volume} {21}},\ \bibinfo {pages} {1487} (\bibinfo
  {year} {2011})}\BibitemShut {NoStop}%
\bibitem [{\citenamefont {Leong}\ \emph {et~al.}(2014)\citenamefont {Leong},
  \citenamefont {Dawson}, \citenamefont {Wirth}, \citenamefont {Li},
  \citenamefont {Connolly}, \citenamefont {Smith}, \citenamefont {Wilkinson},\
  and\ \citenamefont {Miller}}]{Leong2014}%
  \BibitemOpen
  \bibfield  {author} {\bibinfo {author} {\bibfnamefont {H.~S.}\ \bibnamefont
  {Leong}}, \bibinfo {author} {\bibfnamefont {K.}~\bibnamefont {Dawson}},
  \bibinfo {author} {\bibfnamefont {C.}~\bibnamefont {Wirth}}, \bibinfo
  {author} {\bibfnamefont {Y.}~\bibnamefont {Li}}, \bibinfo {author}
  {\bibfnamefont {Y.}~\bibnamefont {Connolly}}, \bibinfo {author}
  {\bibfnamefont {D.~L.}\ \bibnamefont {Smith}}, \bibinfo {author}
  {\bibfnamefont {C.~R.}\ \bibnamefont {Wilkinson}}, \ and\ \bibinfo {author}
  {\bibfnamefont {C.~J.}\ \bibnamefont {Miller}},\ }\href@noop {} {\bibfield
  {journal} {\bibinfo  {journal} {Nat. Commun.}\ }\textbf {\bibinfo {volume}
  {5}},\ \bibinfo {pages} {3947} (\bibinfo {year} {2014})}\BibitemShut
  {NoStop}%
\bibitem [{\citenamefont {Chevin}(2011)}]{Chevin2011}%
  \BibitemOpen
  \bibfield  {author} {\bibinfo {author} {\bibfnamefont {L.-M.}\ \bibnamefont
  {Chevin}},\ }\href@noop {} {\bibfield  {journal} {\bibinfo  {journal} {Biol.
  Lett.}\ }\textbf {\bibinfo {volume} {7}},\ \bibinfo {pages} {210} (\bibinfo
  {year} {2011})}\BibitemShut {NoStop}%
\bibitem [{\citenamefont {Ricklefs}(2010)}]{Ricklefs2010}%
  \BibitemOpen
  \bibfield  {author} {\bibinfo {author} {\bibfnamefont {R.~E.}\ \bibnamefont
  {Ricklefs}},\ }\href@noop {} {\bibfield  {journal} {\bibinfo  {journal}
  {Proc. Natl. Acad. Sci.}\ }\textbf {\bibinfo {volume} {107}},\ \bibinfo
  {pages} {10314} (\bibinfo {year} {2010})}\BibitemShut {NoStop}%
\bibitem [{\citenamefont {Wides}\ and\ \citenamefont {Milo}(2018)}]{Wides2018}%
  \BibitemOpen
  \bibfield  {author} {\bibinfo {author} {\bibfnamefont {A.}~\bibnamefont
  {Wides}}\ and\ \bibinfo {author} {\bibfnamefont {R.}~\bibnamefont {Milo}},\
  }\href@noop {} {\enquote {\bibinfo {title} {Understanding the dynamics and
  optimizing the performance of chemostat selection experiments},}\ }\bibinfo
  {howpublished} {arXiv preprint arXiv:1806.00272} (\bibinfo {year}
  {2018})\BibitemShut {NoStop}%
\bibitem [{\citenamefont {Milo}\ and\ \citenamefont {Phillips}(2017)}]{bionum}%
  \BibitemOpen
  \bibfield  {author} {\bibinfo {author} {\bibfnamefont {R.}~\bibnamefont
  {Milo}}\ and\ \bibinfo {author} {\bibfnamefont {R.}~\bibnamefont
  {Phillips}},\ }\href@noop {} {\enquote {\bibinfo {title} {Cell biology by the
  numbers},}\ }\bibinfo {howpublished} {\url{http://book.bionumbers.org/}}
  (\bibinfo {year} {2017})\BibitemShut {NoStop}%
\bibitem [{\citenamefont {Nei}(2005)}]{nei2005selectionism}%
  \BibitemOpen
  \bibfield  {author} {\bibinfo {author} {\bibfnamefont {M.}~\bibnamefont
  {Nei}},\ }\href@noop {} {\bibfield  {journal} {\bibinfo  {journal} {Molecular
  biology and evolution}\ }\textbf {\bibinfo {volume} {22}},\ \bibinfo {pages}
  {2318} (\bibinfo {year} {2005})}\BibitemShut {NoStop}%
\bibitem [{\citenamefont {Tokuyama}\ and\ \citenamefont
  {Oppenheim}(1995)}]{tokuyama1995theory}%
  \BibitemOpen
  \bibfield  {author} {\bibinfo {author} {\bibfnamefont {M.}~\bibnamefont
  {Tokuyama}}\ and\ \bibinfo {author} {\bibfnamefont {I.}~\bibnamefont
  {Oppenheim}},\ }\href@noop {} {\bibfield  {journal} {\bibinfo  {journal}
  {Physica A}\ }\textbf {\bibinfo {volume} {216}},\ \bibinfo {pages} {85}
  (\bibinfo {year} {1995})}\BibitemShut {NoStop}%
\end{thebibliography}
\end{document}